\documentclass[aps,prb,twocolumn,superscriptaddress,amsmath,amssymb,titlepage]{revtex4-2}
\usepackage{hyperref}
\usepackage{graphicx}
\usepackage{dcolumn}
\usepackage{bm}
\usepackage{tikz}
\usepackage{dsfont}
\usepackage{subfigure}
\usepackage{braket,slashed}
\usepackage{cleveref}

\newenvironment{diagram}
{
\begin{tikzpicture}[baseline = (X.base),every node/.style={scale=0.7},scale=.55]
}
{
\end{tikzpicture}
}

\newcommand{\e}{\ensuremath{\mathrm{e}}}

\begin{document}

\title{Detecting emergent continuous symmetries at quantum criticality}

\author{Mingru Yang}
\affiliation{University of Vienna, Faculty of Physics, Boltzmanngasse 5, 1090 Wien, Austria}
\author{Bram Vanhecke}
\affiliation{University of Vienna, Faculty of Physics, Boltzmanngasse 5, 1090 Wien, Austria}
\author{Norbert Schuch}
\affiliation{University of Vienna, Faculty of Physics, Boltzmanngasse 5, 1090 Wien, Austria}
\affiliation{University of Vienna, Faculty of Mathematics, Oskar-Morgenstern-Platz 1, 1090 Wien, Austria}

\date{\today}

\begin{abstract}
New or enlarged symmetries can emerge at the low-energy spectrum of a Hamiltonian that does not possess the symmetries, if the symmetry breaking terms in the Hamiltonian are irrelevant under the renormalization group flow. In this letter, we propose a tensor network based algorithm to numerically extract lattice operator approximation of the emergent conserved currents from the ground state of any quantum spin chains, without the necessity to have prior knowledge about its low-energy effective field theory. Our results for the spin-1/2 $J$-$Q$ Heisenberg chain and a one-dimensional version of the deconfined quantum critical points (DQCP) demonstrate the power of our method to obtain the emergent lattice Kac-Moody generators. It can also be viewed as a way to find the local integrals of motion of an integrable model and the local parent Hamiltonian of a critical gapless ground state.
\end{abstract}

\maketitle

\emph{Introduction.}---Low-energy physics can show different symmetries from the Hamiltonian. In the thermodynamic limit, the continuous symmetry of a Hamiltonian can be spontaneously broken in its ground state, or new symmetries that the Hamiltonian does not possess can \emph{emerge} in its low-energy spectrum. The latter phenomenon of emergent symmetries is prevalent at the critical point of many quantum and classical phase transitions, provided the symmetry breaking terms in the Hamiltonian are irrelevant under the renormalization group (RG) flow. The most prominent example might be the deconfined quantum critical point (DQCP)~\cite{doi:10.1126/science.1091806,PhysRevB.70.144407}, a direct continuous phase transition between two distinct spontaneous symmetry broken phases without fine-tuning, beyond the Landau-Ginzburg-Wilson paradigm. The emergent symmetry which reconciles the incompatible order parameters thus becomes the smoking gun to determine whether such a phase transition is really a DQCP. Another example is the extended symmetry in the low-energy eigenstates of a one-dimensional (1D) critical Hamiltonian with an internal semi-simple Lie group symmetry, when its low-energy physics is described by a conformal field theory (CFT)~\cite{cft,appliedcft}. In this case, the microscopic symmetry and the emergent symmetries can be recombined to form two independent symmetries acting respectively on the left- and right-moving fields, with the corresponding conserved charges being the zero modes of the Kac-Moody algebra~\cite{PhysRevLett.55.1355,PhysRevB.106.115111}.

Plenty of numerical efforts~\cite{PhysRevB.98.014414,PhysRevB.100.125137,PhysRevLett.122.175701,https://doi.org/10.48550/arxiv.2008.04833} have been devoted to confirming the existence of emergent symmetries. In the case of DQCP, the identity between the scaling dimensions of the critical fluctuations related by emergent symmetries would be an indication~\cite{PhysRevLett.98.227202,PhysRevB.100.125137}. Other approaches include order parameter histograms~\cite{PhysRevLett.98.227202,lee2022landauforbidden} and level-crossing analysis~\cite{PhysRevLett.121.107202}. A more direct probe of emergent symmetries is to check if the scaling dimensions of the effective lattice operators for the conserved currents in the field theory are equal to the space dimension~\cite{PhysRevLett.122.175701,PhysRevB.100.125137}. However, identification of lattice operators to the currents in the continuum limit requires involved field theory and symmetry analysis~\cite{PhysRevB.72.104404,PhysRevB.100.125137}. Moreover, the identification is usually only approximate and also not unique.

Instead, tensor networks~\cite{PhysRevLett.69.2863,PhysRevB.48.10345,Orus2019,RevModPhys.93.045003} provide us with much more information than simply a measurement outcome of the correlation function for given operators. In fact, rather than derive from field theory analysis, we are able to read out the lattice operator for the emergent conserved currents from a tensor network state in a straightforward way. Upon feeding a variationally optimized tensor network ground state~\cite{PhysRevB.97.045145,10.21468/SciPostPhysLectNotes.7},
our algorithm returns the optimal lattice approximation of the conserved current operators truncated to a given interaction range $N$, which systematically approximates the exact symmetry generators as $N$ increases.

\emph{Algorithm.}---If a state $|\psi\rangle$ is symmetric under a global continuous symmetry transformation $U=e^{\mathrm{i}\epsilon O}$, then $U|\psi\rangle=e^{\mathrm{i}\epsilon\phi}|\psi\rangle$. After absorbing the phase factor into the definition of $O$, i.e. $O\rightarrow O-\phi I$, we have $e^{\mathrm{i}\epsilon O}|\psi\rangle=|\psi\rangle$, and its linearization gives
\begin{equation}
O|\psi\rangle=0,
\end{equation}
or $\langle\psi | O^{\dag} O| \psi\rangle=0$. For an internal symmetry with local generators, $O=\sum_n e^{\mathrm{i}pn}G_{n,\dots,n+N-1}$, where $p$ is the momentum and $G_{n,\dots,n+N-1}$ is a $N$-site operator starting at the $n$th site. Given a state $|\psi\rangle$ and a momentum $p$, if we aim to obtain an exact or approximate conserved quantity of this form which the state has, we can consider the optimization problem 
\begin{equation}
\min_{G} f(G,G^{\dag})=\min_{G}\frac{\langle\psi|O^{\dag}O|\psi\rangle}{V\,\mathrm{Tr}[G^{\dag}G]}
\end{equation}
with the normalization constraint $\|G\|^2=\mathrm{Tr}[G^{\dag}G]=1$, where $V$ is the system size. Note that this cost function has a physical interpretation of the static structure factor of $G$ at momentum $p$. The unitarity of $U$ requires $O$, and thus $G$, to be Hermitian. In that case, the optimum of $f$ is reached when $\frac{\partial f}{\partial G}=0$, i.e.
\begin{equation}
\langle\psi|\frac{\partial O^{\dag}}{\partial G} O|\psi\rangle+\langle\psi|O^{\dag}\frac{\partial O}{\partial G}|\psi\rangle=2\frac{\langle\psi|O^{\dag}O|\psi\rangle}{\mathrm{Tr}[G^2]} G,  
\end{equation}
which, after vectorizing $G\mapsto \mathbf{g}$, becomes an eigenvalue problem
\begin{equation}\label{eqn:eigproblem}
    (\mathcal{F}+\mathcal{F}^T)\cdot\mathbf{g} =2\lambda_{min}\mathbf{g},
\end{equation}
where the eigenvalues are guaranteed to be non-negative real numbers due to the positive semi-definite quadratic form of the cost function $f$, and it can be proved~\cite{supp} that the eigenvectors $G$ are guaranteed to be Hermitian up to an arbitrary overall phase. For an eigenvector $G$, the associated eigenvalue $\lambda$ naturally measures how accurate the corresponding symmetry is. 

For an infinite matrix product state (MPS) $|\psi\rangle$, this eigenvalue problem can be solved by adapting MPS techniques used in other contexts; readers not interested in these details can skip this paragraph. Take $|\psi\rangle$ as an infinite uniform MPS with one-site unit cell parameterized by tensors $A_L$, $A_R$, and $A_C$ in the mixed gauge. The application of $\mathcal{F}$ to $\mathbf{g}$, and similarly $\mathcal{F}^T\cdot \mathbf{g}$, can be implemented by observing that~\footnote{We adopt the same notation as in Ref.~\cite{10.21468/SciPostPhysLectNotes.7}} it is the same as calculating the static structure factor of $G$ except that a hole is dug in all the terms, i.e.
\begin{widetext}
\begin{equation}\label{eqn:Fg}
\begin{split}
\mathcal{F}\cdot\mathbf{g} &=\langle\psi |\frac{\partial O^{\dag}}{\partial G}O|\psi\rangle\\
&=\langle\psi |\left(
\frac{1}{V}\sum_m e^{-\mathrm{i}pm}
 \dots~~
\scalebox{0.8}{
\begin{diagram}
\draw (3.5,0.1) node {$m$};
\draw (5.5,0.1) node {$m+N-1$};
\draw (4.5,1.5) node (X) {};
\draw (4.5,0.75) node {$\cdots$};
\draw (4.5,2.25) node {$\cdots$};
\draw (1.5,2.5) -- (1.5,.5); \draw (2.5,2.5) -- (2.5,.5); 
\draw (3.5,1) -- (3.5,.5); \draw (3.5,2) -- (3.5,2.5); \draw (5.5,2) -- (5.5,2.5); \draw (5.5,1) -- (5.5,.5);
\draw (6.5,2.5) -- (6.5,.5); \draw (7.5,2.5) -- (7.5,.5);
\end{diagram}
} ~~\dots
\right)\left(
\sum_n e^{\mathrm{i}pn}
 \dots~~
\scalebox{0.8}{
\begin{diagram}
\draw[rounded corners] (3,2) rectangle (6,1);
\draw (4.5,1.5) node (X) {$G$};
\draw (3.5,0.1) node {$n$};
\draw (5.5,0.1) node {$n+N-1$};
\draw (4.5,0.75) node {$\cdots$};
\draw (4.5,2.25) node {$\cdots$};
\draw (1.5,2.5) -- (1.5,.5); \draw (2.5,2.5) -- (2.5,.5); 
\draw (3.5,1) -- (3.5,.5); \draw (3.5,2) -- (3.5,2.5); \draw (5.5,2) -- (5.5,2.5); \draw (5.5,1) -- (5.5,.5);
\draw (6.5,2.5) -- (6.5,.5); \draw (7.5,2.5) -- (7.5,.5);
\end{diagram}
}~~\dots
\right)|\psi\rangle\\
&= e^{-\mathrm{i}pN}
\scalebox{0.8}{
\begin{diagram}
\draw (1,0) edge[out=180,in=180] (1,3);
\draw[rounded corners] (1,3.5) rectangle (2,2.5);
\draw[rounded corners] (1,0.5) rectangle (2,-0.5);
\draw[rounded corners] (1,1) rectangle (4,2);
\draw (2.5,1.5) node {$G$};
\draw (2.5,0.75) node {$\cdots$};
\draw (2.5,2.25) node {$\cdots$};
\draw (1.5,3) node {$A_L$};
\draw (1.5,0) node {$\bar{A}_L$};
\draw (1.5,2.5) -- (1.5,2);
\draw (1.5,0.5) -- (1.5,1);
\draw (2,3) -- (3,3); \draw (2,0) -- (3,0);
\draw[rounded corners] (3,3.5) rectangle (4,2.5);
\draw[rounded corners] (3,0.5) rectangle (4,-0.5);
\draw (3.5,3) node {$A_L$};
\draw (3.5,0) node {$\bar{A}_L$};
\draw (3.5,2.5) -- (3.5,2);
\draw (3.5,0.5) -- (3.5,1);
\draw (4,3) -- (5,3); \draw (4,0) -- (5,0);
\draw[rounded corners] (5,3.5) rectangle (9,-0.5);
\draw (7,1.5) node {$\left( 1 -  e^{-\mathrm{i}p}E^L_L \right)^{P}$};
\draw (9,3) -- (10,3);  \draw (9,0) -- (10,0);
\draw[rounded corners] (10,3.5) rectangle (11,2.5);
\draw[rounded corners] (10,0.5) rectangle (11,-0.5);
\draw (10.5,3) node {$A_L$};
\draw (10.5,0) node {$\bar{A}_L$};
\draw (10.5,2.5) -- (10.5,2);
\draw (10.5,0.5) -- (10.5,1);
\draw (11,3) -- (12,3);  \draw (11,0) -- (12,0);
\draw[rounded corners] (12,3.5) rectangle (13,2.5);
\draw[rounded corners] (12,0.5) rectangle (13,-0.5);
\draw (12.5,3) node {$A_C$};
\draw (12.5,0) node {$\bar{A}_C$};
\draw (12.5,2.5) -- (12.5,2);
\draw (12.5,0.5) -- (12.5,1);
\draw (11.5,0.75) node {$\cdots$};
\draw (11.5,2.25) node {$\cdots$};
\draw (13,3) edge[out=0,in=0] (13,0);
\end{diagram}
}
+e^{\mathrm{i}pN}
\scalebox{0.8}{
\begin{diagram}
\draw (1,0) edge[out=180,in=180] (1,3);
\draw (2.5,0.75) node {$\cdots$};
\draw (2.5,2.25) node {$\cdots$};
\draw[rounded corners] (1,3.5) rectangle (2,2.5);
\draw[rounded corners] (1,0.5) rectangle (2,-0.5);
\draw (1.5,3) node {$A_C$};
\draw (1.5,0) node {$\bar{A}_C$};
\draw (1.5,2.5) -- (1.5,2);
\draw (1.5,0.5) -- (1.5,1);
\draw (2,3) -- (3,3); \draw (2,0) -- (3,0);
\draw[rounded corners] (3,3.5) rectangle (4,2.5);
\draw[rounded corners] (3,0.5) rectangle (4,-0.5);
\draw (3.5,3) node {$A_R$};
\draw (3.5,0) node {$\bar{A}_R$};
\draw (3.5,2.5) -- (3.5,2);
\draw (3.5,0.5) -- (3.5,1);
\draw (4,3) -- (5,3); \draw (4,0) -- (5,0);
\draw[rounded corners] (5,3.5) rectangle (9,-0.5);
\draw (7,1.5) node {$\left( 1 - e^{\mathrm{i}p}E^R_R \right)^{P}$};
\draw (9,3) -- (10,3);  \draw (9,0) -- (10,0);
\draw[rounded corners] (10,3.5) rectangle (11,2.5);
\draw[rounded corners] (10,0.5) rectangle (11,-0.5);
\draw (10.5,3) node {$A_R$};
\draw (10.5,0) node {$\bar{A}_R$};
\draw[rounded corners] (10,1) rectangle (13,2);
\draw (11.5,1.5) node {$G$};
\draw (11.5,0.75) node {$\cdots$};
\draw (11.5,2.25) node {$\cdots$};
\draw (10.5,2.5) -- (10.5,2);
\draw (10.5,0.5) -- (10.5,1);
\draw (11,3) -- (12,3);  \draw (11,0) -- (12,0);
\draw[rounded corners] (12,3.5) rectangle (13,2.5);
\draw[rounded corners] (12,0.5) rectangle (13,-0.5);
\draw (12.5,3) node {$A_R$};
\draw (12.5,0) node {$\bar{A}_R$};
\draw (12.5,2.5) -- (12.5,2);
\draw (12.5,0.5) -- (12.5,1);
\draw (13,3) edge[out=0,in=0] (13,0);
\end{diagram}
}\;\\
&+ e^{-\mathrm{i}p(N-1)}
\scalebox{0.8}{
\begin{diagram}
\draw (-.5,-1) edge[out=180,in=180] (-.5,4);
\draw[rounded corners] (-0.5,4.5) rectangle (+0.5,3.5);
\draw[rounded corners] (-0.5,-1.5) rectangle (+0.5,-.5);
\draw (.5,4) -- (1.5,4); \draw (.5,-1) -- (1.5,-1);
\draw[rounded corners] (1.5,4.5) rectangle (2.5,3.5);
\draw[rounded corners] (1.5,-1.5) rectangle (2.5,-.5);
\draw (2.5,4) -- (3.5,4); \draw (2.5,-1) -- (3.5,-1);
\draw[rounded corners] (3.5,4.5) rectangle (4.5,3.5);
\draw[rounded corners] (3.5,-1.5) rectangle (4.5,-.5);
\draw[rounded corners] (-.5,2.75) rectangle (2.5,1.75);
\draw (1,2.25) node {$G$};
\draw (1,3.25) node {$\cdots$};
\draw (1,0) node {$\cdots$};
\draw (3,.75) node {$\cdots$};
\draw (0,3.5) -- (0,2.75); \draw (0,1.75) -- (0,-.5);
\draw (2,3.5) -- (2,2.75); \draw (2,1.75) -- (2,1.25); \draw (2,.25) -- (2,-.5);
\draw (4,3.5) -- (4,1.25); \draw (4,.25) -- (4,-.5);
\draw (0,4) node {$A_L$};
\draw (0,-1) node {$\bar{A}_L$};
\draw (2,4) node {$A_L$};
\draw (2,-1) node {$\bar{A}_L$};
\draw (4,4) node {$A_C$};
\draw (4,-1) node {$\bar{A}_C$};
\draw (4.5,4) edge[out=0,in=0] (4.5,-1);
\end{diagram}
}
+ e^{\mathrm{i}p(N-1)}
\scalebox{0.8}{
\begin{diagram}
\draw (-.5,-1) edge[out=180,in=180] (-.5,4);
\draw[rounded corners] (-0.5,4.5) rectangle (+0.5,3.5);
\draw[rounded corners] (-0.5,-1.5) rectangle (+0.5,-.5);
\draw (.5,4) -- (1.5,4); \draw (.5,-1) -- (1.5,-1);
\draw[rounded corners] (1.5,4.5) rectangle (2.5,3.5);
\draw[rounded corners] (1.5,-1.5) rectangle (2.5,-.5);
\draw (2.5,4) -- (3.5,4); \draw (2.5,-1) -- (3.5,-1);
\draw[rounded corners] (3.5,4.5) rectangle (4.5,3.5);
\draw[rounded corners] (3.5,-1.5) rectangle (4.5,-.5);
\draw[rounded corners] (1.5,2.75) rectangle (4.5,1.75);
\draw (3,2.25) node {$G$};
\draw (3,3.25) node {$\cdots$};
\draw (3,0) node {$\cdots$};
\draw (1,.75) node {$\cdots$};
\draw (0,3.5) -- (0,1.25); \draw (0,.25) -- (0,-.5);
\draw (2,3.5) -- (2,2.75); \draw (2,1.75) -- (2,1.25); \draw (2,.25) -- (2,-.5);
\draw (4,3.5) -- (4,2.75); \draw (4,1.75) -- (4,-.5);
\draw (0,4) node {$A_C$};
\draw (0,-1) node {$\bar{A}_C$};
\draw (2,4) node {$A_R$};
\draw (2,-1) node {$\bar{A}_R$};
\draw (4,4) node {$A_R$};
\draw (4,-1) node {$\bar{A}_R$};
\draw (4.5,4) edge[out=0,in=0] (4.5,-1);
\end{diagram}
}
+\dots+
\scalebox{0.8}{
\begin{diagram}
\draw (-.5,-1) edge[out=180,in=180] (-.5,4);
\draw[rounded corners] (-0.5,4.5) rectangle (+0.5,3.5);
\draw[rounded corners] (-0.5,-1.5) rectangle (+0.5,-.5);
\draw (.5,4) -- (1.5,4); \draw (.5,-1) -- (1.5,-1);
\draw[rounded corners] (1.5,4.5) rectangle (2.5,3.5);
\draw[rounded corners] (1.5,-1.5) rectangle (2.5,-.5);
\draw[rounded corners] (-.5,2.75) rectangle (2.5,1.75);
\draw (1,2.25) node {$G$};
\draw (1,3.25) node {$\cdots$};
\draw (1,0) node {$\cdots$};
\draw (1,.75) node {$\cdots$};
\draw (0,3.5) -- (0,2.75); \draw (0,1.75) -- (0,1.25); \draw (0,.25) -- (0,-.5);
\draw (2,3.5) -- (2,2.75); \draw (2,1.75) -- (2,1.25); \draw (2,.25) -- (2,-.5);
\draw (0,4) node {$A_C$};
\draw (0,-1) node {$\bar{A}_C$};
\draw (2,4) node {$A_R$};
\draw (2,-1) node {$\bar{A}_R$};
\draw (2.5,4) edge[out=0,in=0] (2.5,-1);
\end{diagram}
},
\end{split}
\end{equation}
\end{widetext}
where the last ``$\dots$'' means sum over all diagrams with $1\leq |n-m|\leq N-2$, $E^L_L$ and $E^R_R$ are the left- and right-gauge MPS transfer matrices, and $(\cdot)^P$ denotes the pseudo-inverse resulting from the infinite geometric series~\cite{10.21468/SciPostPhysLectNotes.7} of all relative positions between $G$ and the hole without overlap, which includes a regularization procedure effectively removing the disconnected part of the correlation functions and thus is automatically consistent with the phase factor absorption mentioned previously. We can then use an iterative eigensolver~\cite{doi:10.1137/1.9781611970739} to obtain the lowest several solutions~\footnote{Notice that there always exist trivial solutions of the form $G=X\otimes I-I\otimes X$ for $p=0$ and $G=X\otimes I+I\otimes X$ for $p=\pi$, with $X$ being any $N-1$-site operator (except the identity) and $I$ being the one-site identity. The $N$-site identity is also a trivial solution by regularization. The eigenvalues associated with those trivial solutions are exactly zero. In total they span a large trivial null space of dimension $d^{2(N-1)}$, where $d$ is the dimension of the one-site physical Hilbert space. We have removed the trivial solutions from all of our results. Interested readers can go to the supplementary material~\cite{supp} for how to remove the trivial solutions.}.

In principle, the algorithm works for any MPS~\footnote{Solving the symmetry of the ground state of a gapped model is much easier, and another method based on the MPS fundamental theorem also works~\cite{supp}.}. Particularly, we are interested in applying it to the variational uniform MPS (VUMPS)~\cite{PhysRevB.97.045145} approximation of the gapless ground state of 1D critical Hamiltonians. Since an MPS with finite bond dimension is always gapped~\cite{10.21468/SciPostPhysLectNotes.7}, it can never exactly represent a critical ground state of infinite correlation length and thus can never exactly capture the symmetry of a critical lattice Hamiltonian or of its low-energy effective field theory in the infrared limit. However, we may use the principle of entanglement scaling~\cite{PhysRevB.78.024410,PhysRevLett.102.255701,PhysRevX.8.041033} and treat the finite bond dimension $\chi$ as a relevant perturbation, which enables us to identify the exact or emergent symmetries exclusively from the MPS through an extrapolation in the correlation length $\xi$, as shown by the benchmark results below.

\emph{Benchmarks for exact symmetries.}%
---As a warming up, we first consider a critical model whose ground state has an exact $\mathrm{U}(1)$ symmetry~\footnote{In fact the symmetry should be $\mathrm{U}(1)\times\mathrm{U}(1)$. See Ref.~\cite{PhysRevLett.55.1355}.}---the spin-1/2 isotropic quantum $XY$ chain
\begin{equation}
\label{Eq:HXY}
H = -\sum_n\left(X_nX_{n+1}+Y_nY_{n+1}\right),
\end{equation}
where $X_n$, $Y_n$, and $Z_n$ are the Pauli matrices at site $n$. The $\mathrm{U}(1)$ symmetry is generated by $O=\sum_nZ_n$ that satisfies $[H,\sum_nZ_n]$=0. The model is integrable and thus has infinitely many local conserved quantities in the thermodynamic limit~\cite{https://doi.org/10.1002/sapm1985733221,Araki1990,Matsui1993,Fagotti_2016}. The critical low-energy physics is described~\cite{PhysRevLett.119.261603} by the $\mathrm{U}(1)_4$ CFT of free bosons with central charge $c=1$.

\begin{figure}
    \includegraphics[scale=0.327]{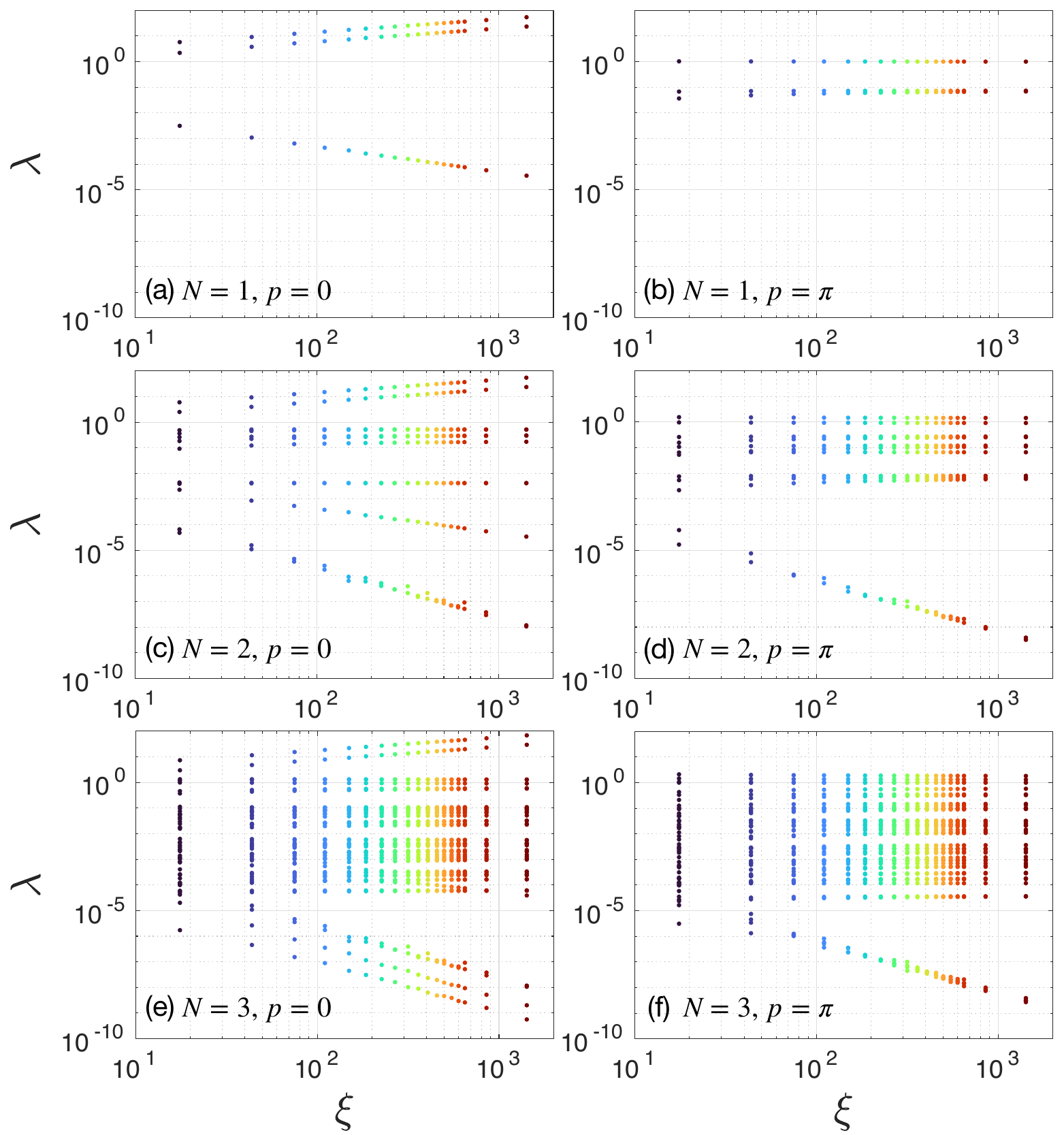}
    \caption{Log-log plot of the eigenvalue spectrum of $\frac{1}{2}(\mathcal{F}+\mathcal{F}^T)$ versus the correlation length $\xi$ for the spin-1/2 isotropic quantum $XY$ chain. The correlation length of an MPS with a certain bond dimension is calculated by Eq. (40) in Ref.~\cite{10.21468/SciPostPhysLectNotes.7}. Notice that in (e) there is one decaying $\lambda$ hidden in the bulk of larger eigenvalues but it becomes visible at larger correlation lengths.}
    \label{fig:XYeig}
\end{figure}

Applying our algorithm to MPS of various bond dimensions yields the local conserved quantities up to $N=3$. The full spectrum (after removing the trivial solutions) of the eigenvalue problem in Eq.~(\ref{eqn:eigproblem}) are shown in Fig.~\ref{fig:XYeig} and the eigenvectors $G$ associated with the decaying eigenvalues are shown in Table~\ref{table:XYg}. For $p=0$, there are 1, 3, 5 eigenvalues decaying with the correlation length for $N=1,2,3$, respectively; for $p=\pi$, there are 0, 2, 4 eigenvalues decaying with the correlation length for $N=1,2,3$, respectively. The eigenvector $G=XX+YY$ corresponds to the Hamiltonian in Eq.~(\ref{Eq:HXY}), so as a by-product our method is also able to determine the local parent Hamiltonian~\cite{PhysRevX.8.021026,Qi2019determininglocal,PhysRevX.8.031029} solely from its ground state. We notice that the decay has a power-law scaling $\lambda\sim \xi^{-\eta'}$~\footnote{Depending on the quantum number the operator carries, the correlation length of correlation functions of an operator can be different from the correlation length of the MPS~\cite{supp}. It turns out that the critical exponents here have no universal relation to the scaling dimension of the corresponding operator~\cite{supp}.} and the exponents are listed in Table~\ref{table:XYg}. All other eigenvalues increase or stay constant with the increasing correlation length. The $G$'s associated with the decaying $\lambda$'s are local integrals of motion since $\lambda$ is extrapolated to 0 at infinite correlation length. While the conserved quantities in Table~\ref{table:XYg} and more conserved quantities for larger $N$ in the $XY$ model can be constructed recursively~\cite{supp} through the master symmetry approach~\cite{mastersymmetry,https://doi.org/10.1002/sapm1985733221,Araki1990,Matsui1993,PhysRevB.106.115111}, our method provides an alternative way to obtain them generally.

\begin{table}
\centering
\begin{tabular}{ c|c|c|c }
\hline\hline
 $p$ & $N$ & $G$ & $\eta'$ \\ 
\hline\hline
                     & ~~~~1~~~~            & $Z$      & ~~~~1.009~~~~ \\\cline{2-4}
                     & ~~~~2~~~~            & $XX+YY$  & 1.985 \\
~~~~0~~~~            &                      & $XY-YX$  & 1.933 \\\cline{2-4}
                     & ~~~~3~~~~            & $XZX+YZY$ & 1.008 \\
                     &                      & $XZY-YZX$ & 1.939 \\
\hline
                     & ~~~~1~~~~            & ---      & --- \\\cline{2-4}
                     & ~~~~2~~~~            & $XX-YY$  & 2.005 \\
~~~~$\pi$~~~~        &                      & $XY+YX$  & 2.008 \\\cline{2-4}
                     & ~~~~3~~~~            & $XZX-YZY$ & 2.046 \\
                     &                      & $XZY+YZX$ & 2.063 \\                    
\hline\hline
\end{tabular}
\caption{Local conserved quantities in the spin-1/2 isotropic quantum $XY$ chain up to $N=3$. Smaller-$N$ solutions reappear at larger $N$ and we only show the new solutions at each $N$. The $\eta'$ is obtained from the scaling of $\langle\psi |O^{\dag}O|\psi\rangle$ with $\xi$, which is slightly different from the slope of the decaying eigenvalues in Fig. \ref{fig:XYeig}, since different solutions can mix with each other and their form also become more accurate as $\xi$ increases.}
\label{table:XYg}
\end{table}

\emph{Extended symmetries by emergent symmetries.}%
---The ground state of the spin-1/2 antiferromagnetic Heisenberg chain is expected to have an emergent symmetry in addition to the microscopic $\mathrm{SU}(2)$ symmetry of the lattice Hamiltonian, and thus the symmetry is extended to $\mathrm{SO(4)}=[\mathrm{SU(2)}_L\times\mathrm{SU(2)}_R]/\mathbb{Z}_2$~\cite{PhysRevLett.55.1355,PhysRevB.36.5291}. Here, we consider the $J$-$Q$ model~\cite{PhysRevB.98.014414}---a modified Heisenberg chain at whose transition point still exists the extended symmetry---
\begin{equation}
\label{eq:HJQ}
H = -J\sum_nP_{n,n+1}-Q\sum_nP_{n,n+1}P_{n+2,n+3},
\end{equation}
where $P_{n,n+1}=1/4-\mathbf{S}_n\cdot\mathbf{S}_{n+1}$ with $\mathbf{S}_n=(S_n^x,S_n^y,S_n^z)=\frac{1}{2}(X_n,Y_n,Z_n)$. The dimer order enforced by strong four-site interaction transits to a critical phase when $Q/J\lesssim 0.84831$~\cite{PhysRevLett.107.157201,PhysRevB.84.235129}, at which~\footnote{The marginally irrelevant spin-umklapp term which introduces a logarithmic correction to the correlation function vanishes at the transition point, like in the $J_1$-$J_2$ model~\cite{PhysRevLett.60.635,PhysRevB.98.014414}.} the effective description is the $c=1$ $\mathrm{SU}(2)_1$ Wess-Zumino-Witten (WZW) CFT~\cite{PhysRevLett.55.1355,PhysRevB.36.5291}.

\begin{figure}
    \includegraphics[scale=0.312]{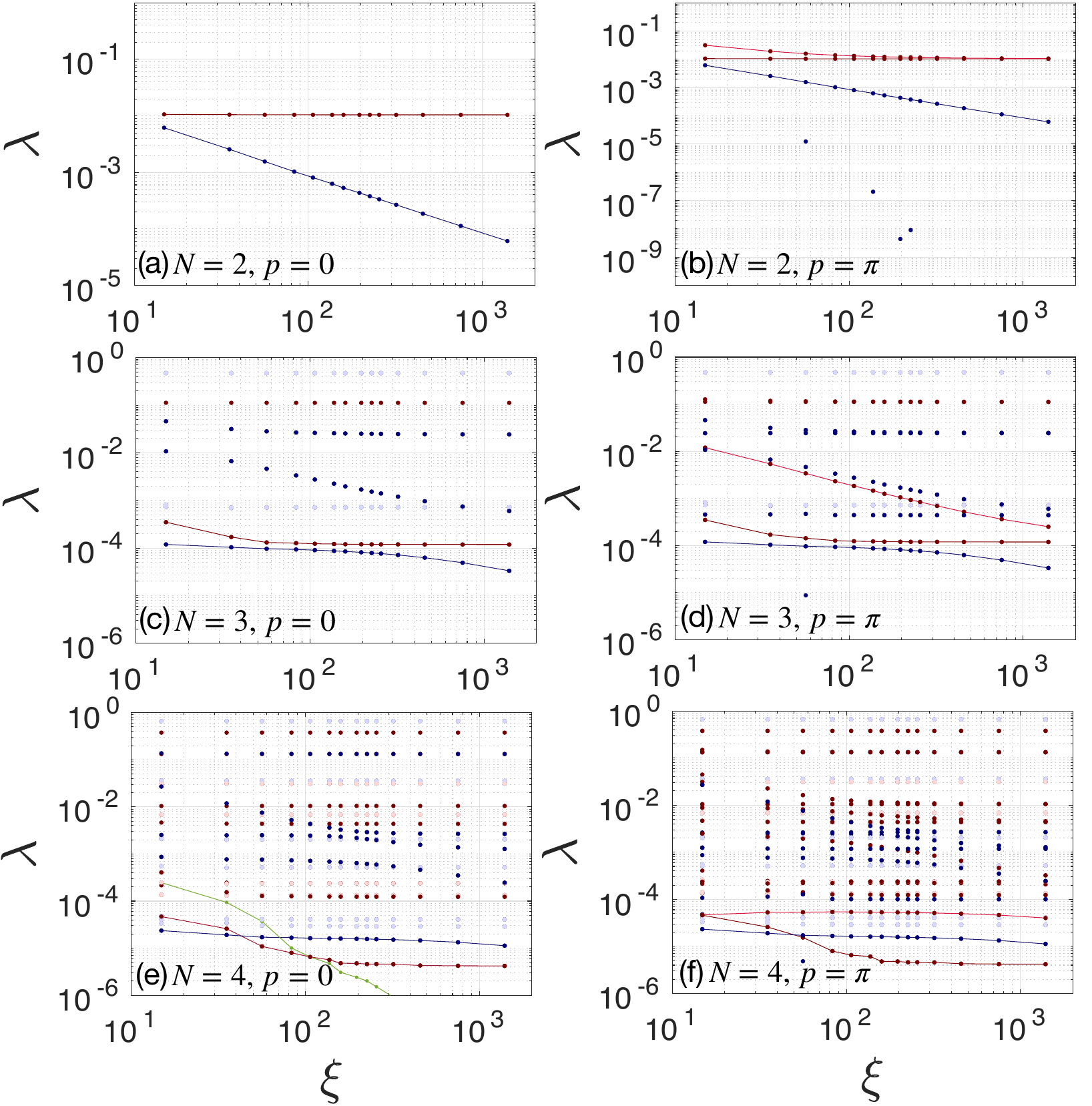}
    \caption{Eigenvalue spectrum for the spin-1/2 $J$-$Q$ Heisenberg model after imposing microscopic symmetries. The $G$'s associated with the blue (red) dots are parity even (odd) and time reversal odd (even). The scaling dimension of the $G$ associated with the faded dots is not one and thus they are not emergent continuous internal symmetries. The green curve in (e) corresponds to the Hamiltonian. Notice that in (d)(f) one solution corresponding to one of the three generators for the exact $\mathrm{SU}(2)$ symmetry of the microscopic Hamiltonian is not shown since it is well below $10^{-6}$. We use a sublattice rotation about the $z$ axis by angle $\pi$ when using VUMPS to optimize the ground state, so the $x$ and $y$ components of the generators move to $p=\pi$.}
    \label{fig:JQeig}
\end{figure}

Fig.~\ref{fig:JQeig} shows the eigenvalues of our optimization problem after imposing the time reversal, parity, and spin flip symmetries~\cite{supp} of the microscopic Hamiltonian at the transition point. The eigenvectors associated with all the eigenvalues shown in Fig.~\ref{fig:JQeig} except the faded ones are lattice operator approximation for the conserved currents of the extended symmetry to different precision, which could be confirmed by checking~\cite{supp} that their scaling dimension is one~\cite{PhysRevLett.55.1355}. To identify the eigenvalues that associate with the same $G$'s at different $\xi$, we search for the eigenvectors at smaller $\xi$'s which have the largest overlap with each of the lowest several eigenvectors at the largest $\xi$ reached, as tracked by the colored lines in Fig.~\ref{fig:JQeig}. The dots connected by blue and red lines at the bottom of the spectrum are the best approximation among all of the solutions. Different from the exact symmetries, eigenvalues corresponding to the emergent symmetries will finally saturate at some correlation length, because it is only a $N$-site truncated approximation of the exact emergent lattice generator.

We observe that three approximately conserved charges (red curves in Fig.~\ref{fig:JQeig}), $M^{\alpha}=\sum_nm_{n}^{\alpha}$ ($\alpha\in\{x,y,z\}$), coming from the emergent symmetries, begin to appear at $N=2$ in addition to the three exact microscopic $\mathrm{SU}(2)$ symmetry generators (blue curves in Fig.~\ref{fig:JQeig}) $Q^{\alpha}=\sum_nS_n^{\alpha}$, and they become more conserved as $N$ increases, which is obvious from the drop of the corresponding eigenvalues. At $N=2$, $m_{n,\alpha}=\epsilon_{\alpha\beta\gamma}S_n^{\beta}S_{n+1}^{\gamma}$ with $\epsilon_{\alpha\beta\gamma}$ the Levi-Civita symbol; at $N=3$, next-nearest neighbor term shows up and we have $m_{n,\alpha}=\epsilon_{\alpha\beta\gamma}(w_1S_n^{\beta}S_{n+1}^{\gamma}+w_2S_n^{\beta}S_{n+2}^{\gamma})$ with $w_2/w_1\approx 0.2253$~\footnote{This form of the 3-site $m_{n,\alpha}$ and the ratio $w_2/w_1$ in the $J$-$Q$ model is very similar to those in the $J_1$-$J_2$ model~\cite{PhysRevB.106.115111}. They flow to the same RG fixed point and they are very similar even on the lattice level. To check this, we have performed exact diagonalization of a system size of 18 sites
and found that the fidelity between their ground states
is 0.9987.}. The form of the 3-site $m_{n,\alpha}$ looks very similar to the level-1 Yangian~\cite{PhysRevLett.69.2021,JCTalstra_1995,Drinfeld:466366}---which are exact conserved quantities of the RG fixed point, the Haldane-Shastry model~\cite{PhysRevLett.60.635,PhysRevLett.60.639}---truncated to the next-nearest neighbor coupling, though with different coupling coefficients~\cite{supp}. When going to $N=4$, the contribution from longer-range coupling in the level-1 Yangian appears with all coupling coefficients modified. Moreover, terms from the level-3 Yangian begin to be involved. We get $m_{n,\alpha}=m^1_{n,\alpha}+m^3_{n,\alpha}$, where $m^1_{n,\alpha}=\epsilon_{\alpha\beta\gamma}(w_1S_n^{\beta}S_{n+1}^{\gamma}+w_2S_n^{\beta}S_{n+2}^{\gamma}+w_3S_n^{\beta}S_{n+3}^{\gamma})$ and $m^3_{n,\alpha}=\epsilon_{\alpha\beta\gamma}\big[u_1S_n^{\beta}S_{n+3}^{\gamma}\mathbf{S}_{n+1}\cdot\mathbf{S}_{n+2}
+u_2\mathbf{S}_{n}\cdot\mathbf{S}_{n+3}S_{n+1}^{\beta}S_{n+2}^{\gamma}
+u_3(S_n^{\beta}S_{n+1}^{\gamma}\mathbf{S}_{n+2}\cdot\mathbf{S}_{n+3}+\mathbf{S}_{n}\cdot\mathbf{S}_{n+1}S_{n+2}^{\beta}S_{n+3}^{\gamma})
+u_4(S_n^{\beta}S_{n+2}^{\gamma}\mathbf{S}_{n+1}\cdot\mathbf{S}_{n+3}+\mathbf{S}_{n}\cdot\mathbf{S}_{n+2}S_{n+1}^{\beta}S_{n+3}^{\gamma})\big]$,
with $w_2/w_1\approx 0.3557$, $w_3/w_1\approx 0.1467$, $u_1/w_1\approx 0.1577$, $u_2/w_1\approx -0.09690$, $u_3/w_1\approx -0.09141$, and $u_4/w_1\approx 0.08169$. Considering that it is even under time reversal and odd under parity, $M^{\alpha}\sim J_0^{\alpha}-\bar{J}_0^{\alpha}$, where $J_0^{\alpha}$ ($\bar{J}_0^{\alpha}$) is the zero mode of the Kac-Moody generators, which form the ordinary Lie algebra $\mathfrak{su}(2)_L$ ($\mathfrak{su}(2)_R$)~\footnote{Notice that $J_{\alpha}+\bar{J}_{\alpha}$ also satisfies the $\mathfrak{su}(2)$ algebra, but $J_{\alpha}-\bar{J}_{\alpha}$ does not.}. Since $Q^{\alpha}\sim J_0^{\alpha}+\bar{J}_0^{\alpha}$~\footnote{Strictly speaking, from bosonization~\cite{PhysRevLett.55.1355} one knows that $S^{\alpha}(x)\sim (J_0^{\alpha}(x)+\bar{J}_0^{\alpha}(x))+(-1)^x$(term with scaling dimension 1/2), however, the latter staggered part would cancel out upon sum over $x$ and we get only $J_0^{\alpha}+\bar{J}_0^{\alpha}$.}, $M^{\alpha}$ and $Q^{\alpha}$ can then be linearly combined to construct $J_0^{\alpha}$ and $\bar{J}_0^{\alpha}$, and other modes of the Kac-Moody generators can be constructed by the Fourier transform of the currents~\cite{PhysRevB.106.115111}.

\begin{figure}
    \includegraphics[scale=0.305]{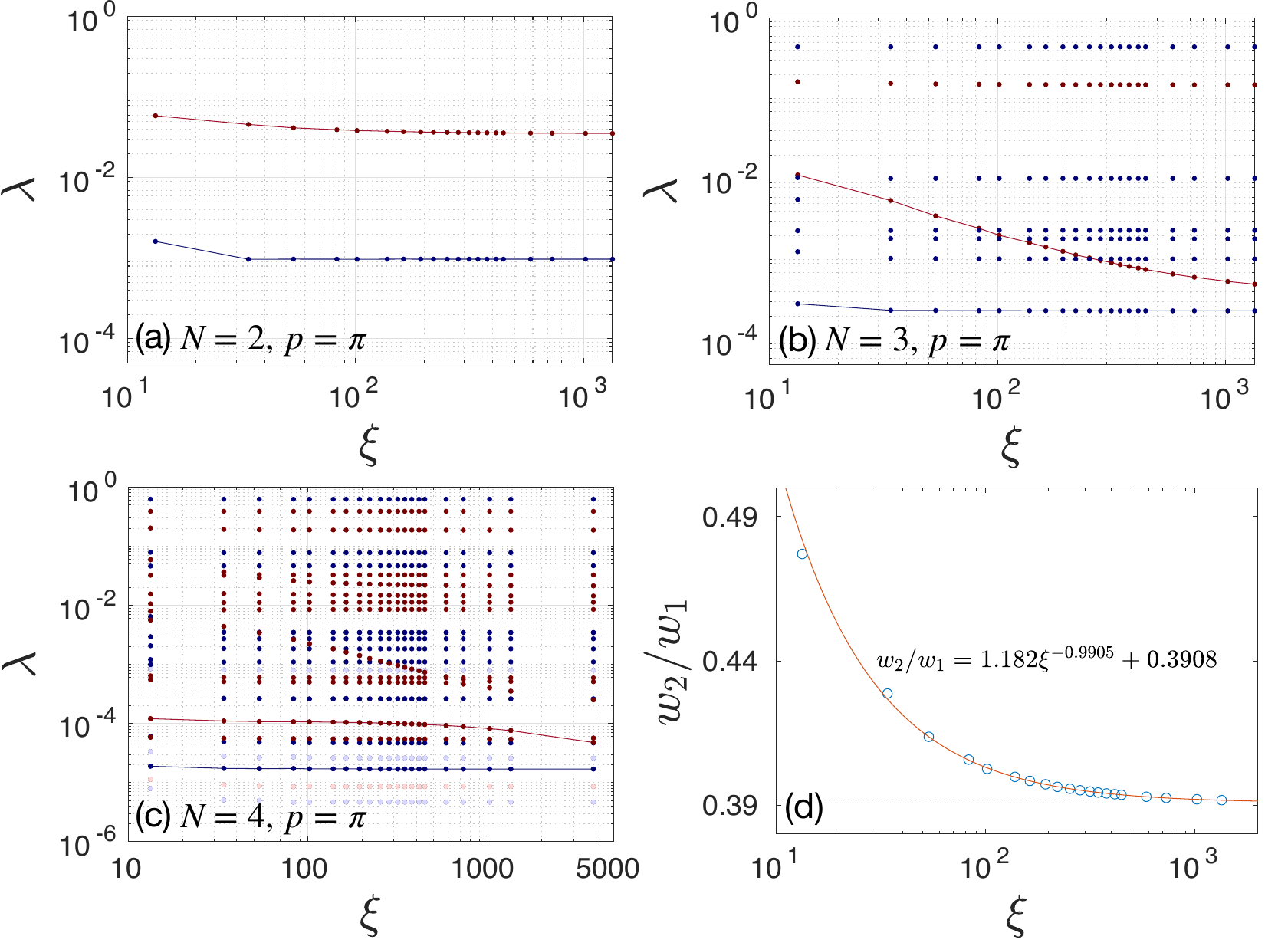}
    \caption{Eigenvalue spectrum for the Jiang-Motrunich model after imposing microscopic symmetries at $p=\pi$ for (a) $N=2$ (b) $N=3$ (c) $N=4$. The color convention is the same as in Fig.~\ref{fig:JQeig}. (d) The ratio between the coefficient in front of the nearest neighbor term and next-nearest neighbor term in $G_2$ at $N=3$ (the red curve in (b)).}
    \label{fig:Motrunicheig}
\end{figure}

\emph{Emergent symmetries at a DQCP.}%
---The following spin-1/2 chain, studied by Jiang and Motrunich~\cite{PhysRevB.99.075103},
\begin{equation}
\begin{split}
H = \sum_n (-J_x&X_nX_{n+1}-J_zZ_nZ_{n+1})\\
&+\sum_n(K_{2x}X_nX_{n+2}+K_{2z}Z_nZ_{n+2}),
\end{split}
\end{equation}
has an onsite $\mathbb{Z}_2\times\mathbb{Z}_2$ spin-flip symmetry. It undergoes a direct continuous transition from a valence bond solid to ferromagnetic order at $K_{2x}=K_{2z}=1/2,~J_x=1,~J_z\approx 1.4645$~\cite{PhysRevB.100.125137}, which has been proposed to be a DQCP with an emergent $\mathrm{U}(1)\times\mathrm{U}(1)$ symmetry~\cite{PhysRevB.99.165143,PhysRevB.100.125137} that is also generated by the zero modes of the Kac-Moody algebra.

Applying our algorithm to the critical point, we find a single solution $G=Z$ for $N=1$. From $N=2$, we require the eigenvectors to transform the same as $Z$ under the spin flip symmetry when solving the eigenvalue problem, and find two solutions at $p=\pi$ as shown in Fig.~\ref{fig:Motrunicheig}(a,b,c).  For $N=2$, the lowest solution (blue) is $G_1=ZI-IZ$ (i.e., a staggered $Z$), and the second solution (red) is $G_2=XY+YX$, which satisfies $[G_1,G_2]=0$; these are indeed precisely the same effective lattice operators identified as conserved currents for the emergent $\mathrm{U}(1)\times\mathrm{U}(1)$ symmetry through bosonization~\cite{PhysRevB.100.125137}. At $N=3$, the corresponding eigenvalues for $G_1$ and $G_2$ improve by almost one and two orders of magnitude, respectively. The form of both solutions modifies significantly by 3-site terms as compared to $N=2$---and thus compared to the field theory prediction---$G_1$ becomes $\frac{v_1}{3}(-ZII+IZI-IIZ)+v_2ZZZ+v_3(YYZ+ZYY)+v_4(XXZ+ZXX)+v_5XZX+v_6YZY$, with $v_2/v_1\approx 0.1615$, $v_3/v_1\approx 0.0988$, $v_4/v_1\approx 0.0882$, $v_5/v_1\approx 0.0410$, and $v_6/v_1\approx -0.1399$; $G_2$ becomes $w_1[(XY+YX)I-I(XY+YX)]+2w_2(XIY-YIX)$, with $w_2/w_1\approx 0.3908$ (Fig.~\ref{fig:Motrunicheig}(d)). When pushing to $N=4$, longer-ranged terms further dress $G_1$ and $G_2$~\cite{supp}. Our algorithm hence allows us to decorate upon the bare form of the lattice operators for emergent symmetry generators found through field theory analysis, and therefore to obtain a more precise picture of the microscopic nature of the emergent symmetries.

\emph{Conclusions.}---We have presented a novel general method to numerically detect emergent continuous internal symmetries in critical systems. The bottom line is that emergent symmetries do not just reveal themselves indirectly in the long-distance behavior of correlation functions---which has been the sole detection mechanism before our work---but are actually realized surprisingly accurately on the lattice, albeit with spatially extended generators. We have illustrated this by rediscovering the theory-predicted lattice operators for the emergent conserved currents at a 1D DQCP and sharply improving them with newly discovered correction terms. We have also identified the effective lattice operators of the conserved charges for the extended $\mathrm{SO}(4)$ symmetry in the \mbox{$J$-$Q$} chain with Yangian generators truncated to local terms, which were unknown before. The ability of our method to crack the explicit form of these lattice generators allows us to construct the emergent lattice Kac-Moody generators to unprecedented accuracy for both Abelian and non-Abelian symmetries~\footnote{Note that it happened to be easy to tell what the symmetry group the lattice generators obey in the above examples, had this not been the case we would have had to study the commutation relations of the generators projected onto the low-energy sector to discover the algebra.} in generic settings.

\emph{Outlook.}---This method could in principle be generalized to 2D, to extract emergent lattice conserved currents in the projected entangled pair states (PEPS)~\cite{https://doi.org/10.48550/arxiv.cond-mat/0407066,PhysRevB.92.201111,PhysRevB.99.165121}, which would be of particular use for the study of higher-dimensional DQCP. A variant version with a larger unit cell can be easily derived. It is also worth exploring if a similar algorithm works for finite systems with periodic or other boundary conditions, for the low-energy excited states~\cite{PhysRevB.85.100408}, or for classical systems. Adjusting this method to find unconventional symmetries of the weakly-entangled higher excited states~\cite{PhysRevLett.111.127201,PhysRevB.107.224312} or the emergent space-time symmetry~\cite{lin2023conformal} would be also interesting directions.

The complexity of the eigenvalue problem scales exponentially with $N$. To reduce the complexity of solving for $G$ of larger size, we could resort to the density matrix renormalization group (DMRG)~\cite{PhysRevLett.69.2863,PhysRevB.48.10345} by treating $G$ as an $N$-site finite matrix product operator (MPO)~\cite{PhysRevA.78.012356,PhysRevB.78.035116}, and the tricky part will be removing the trivial solutions efficiently~\cite{supp,Note2}. It would also be desirable to include terms with long-range tails by representing $O$ as an infinite MPO~\cite{PhysRevB.102.035147}, though its implementation encounters some technical difficulties~\cite{supp}.

\emph{Acknowledgments.}---We thank Frank Verstraete, Natalia Chepiga, Jutho Haegeman, Laurens Vanderstraeten, Wen-Tao Xu, Andr\'as Moln\'ar, Anna Francuz, Ilya Kull, Jos\'e Garre Rubio, Juraj Hasik, Rui-Zhen Huang, Pranay Patil, and especially  Hong-Hao Tu for illuminating discussions. We also acknowledge the hospitality of the Erwin Schr\"odinger International Institute for Mathematics and Physics (ESI) during the long-term program ``Tensor Networks: Mathematical Structures and Novel Algorithms'', which stimulated many of the discussions. This work has received support from the European Union’s Horizon 2020 program through Grant No.\ 863476 (ERC-CoG SEQUAM). The tensor contractions are implemented using \texttt{ncon}~\cite{pfeifer2015ncon}. The codes are available on github~\cite{repo}.

\bibliography{apssamp}
\clearpage

\onecolumngrid
\vspace{\columnsep}
\begin{center}
\textbf{\large Supplementary material for ``Detecting emergent continuous symmetries at quantum criticality''}
\end{center}

\setcounter{equation}{0}
\setcounter{figure}{0}
\setcounter{table}{0}
\pagenumbering{roman}
\setcounter{page}{1}
\makeatletter
\renewcommand{\theequation}{S\arabic{equation}}
\renewcommand{\thefigure}{S\arabic{figure}}
\tableofcontents

\section{Imposing Hermiticity}
In the main text, we mentioned that it can be proved that the eigenvectors $G$ are guaranteed to be Hermitian up to an arbitrary overall phase. In the following, we will explain why this is true.

To make $U=e^{-\mathrm{i}\epsilon O}$ unitary, $O$ must be Hermitian. For $p=0$ or $p=\pi$, $G$ should also be Hermitian. Supposing $G_n$ is a $N$-site Hermitian operator starting at site $n$, then
\begin{equation}
\label{eq:Gansatz}
G_n=\sum_{\alpha=1}^{d^{2N}} c_{\alpha}v_{n,\alpha},
\end{equation}
where $c_{\alpha}$ is real and $v_{\alpha}$ is a $N$-site Kronecker product of the 1-site Hermitian basis operators $\{\frac{1}{\sqrt{2d}}I,T_1,\cdots,T_{d^2-1}\}$, with $d$ the dimension of the 1-site Hilbert space and $T_{i}$ the $\mathrm{SU}(d)$ generator in the fundamental representation.

Let's take $p=0$ as an example. In this case, we have $O=\sum_nG_n$. The cost function will then become
\begin{equation}
f=\frac{\langle\psi|O^{\dag}O|\psi\rangle}{V\,\mathrm{Tr}[G^{\dag}G]}=\frac{\sum_{n,m}\sum_{\alpha,\beta} c_{\alpha}c_{\beta}\langle\psi|v_{m,\beta}v_{n,\alpha}|\psi\rangle}{V\sum_{\alpha,\beta} c_{\alpha}c_{\beta}\mathrm{Tr}(v_{\alpha} v_{\beta})}=\frac{2^N\sum_{n}\sum_{\alpha,\beta} c_{\alpha}c_{\beta}\langle\psi|v_{0,\beta}v_{n,\alpha}|\psi\rangle}{\sum_{\alpha} c_{\alpha}^2}=2^N\frac{\mathbf{c}^{T}\cdot M\cdot\mathbf{c}}{\mathbf{c}^{T}\cdot\mathbf{c}},
\end{equation}
where we have used the fact that $\mathrm{Tr}(v_{\alpha} v_{\beta})=\frac{1}{2^N}\delta_{\alpha\beta}$ and translational invariance, with
\begin{equation}
M_{\beta\alpha}=\sum_n\langle\psi|v_{0,\beta}v_{n,\alpha}|\psi\rangle.
\end{equation}
Then the first derivative becomes
\begin{equation}
\begin{split}
\frac{\partial f}{\partial c_{\gamma}}&=2^N\frac{(\sum_{n}\sum_{\beta}c_{\beta}\langle\psi|v_{0,\beta}v_{n,\gamma}|\psi\rangle+\sum_{n}\sum_{\alpha}c_{\alpha}\langle\psi|v_{0,\gamma}v_{n,\alpha}|\psi\rangle)(\sum_{\alpha} c_{\alpha}^2)-(\sum_{n}\sum_{\alpha,\beta} c_{\alpha}c_{\beta}\langle\psi|v_{0,\beta}v_{n,\alpha}|\psi\rangle)(2c_{\gamma})}{(\sum_{\alpha} c_{\alpha}^2)^2}\\
&=2^N\left[\frac{(\mathbf{c}^T\cdot M)_{\gamma}+(M\cdot\mathbf{c})_{\gamma}}{\mathbf{c}^T\cdot\mathbf{c}}-\frac{2(\mathbf{c}^T\cdot M\cdot\mathbf{c})c_{\gamma}}{(\mathbf{c}^T\cdot\mathbf{c})^2}\right].
\end{split}
\end{equation}
If we let $\frac{\partial f}{\partial c_{\gamma}}=0$, we will get
\begin{equation}
(\mathbf{c}^T\cdot M)_{\gamma}+(M\cdot\mathbf{c})_{\gamma}=\frac{2(\mathbf{c}^T\cdot M\cdot\mathbf{c})}{\mathbf{c}^T\cdot\mathbf{c}}c_{\gamma},
\end{equation}
or
\begin{equation}
[(M^T+M)\cdot\mathbf{c}]_{\gamma}=\frac{\mathbf{c}^T\cdot (M^T+M)\cdot\mathbf{c}}{\mathbf{c}^T\cdot\mathbf{c}}c_{\gamma}.
\end{equation}
Therefore, in order to minimize $f$, we need to solve the eigenvalue problem
\begin{equation}
\label{eq:eigM}
(M^T+M)\cdot\mathbf{c}=2\tilde{\lambda}_{min}\mathbf{c}
\end{equation}
with the constraint that $\mathbf{c}$ is a real vector, which is automatically satisfied. This could be proved as the following. By utilizing translational invariance and the Hermiticity of $v_{n,\alpha}$, we can get that $M$ is Hermitian, i.e.
\begin{equation}
M_{\alpha\beta}^*=\sum_{n}\langle\psi |v_{0,\alpha}v_{n,\beta}|\psi\rangle^*=\sum_{n}\langle\psi |v_{-n,\alpha}v_{0,\beta}|\psi\rangle^*=\sum_{n}\langle\psi |v_{n,\alpha}v_{0,\beta}|\psi\rangle^*=\sum_{n}\langle\psi |v_{0,\beta}v_{n,\alpha}|\psi\rangle=M_{\beta\alpha}.
\end{equation}
So
\begin{equation}
M_{\alpha\beta}^*+M_{\beta\alpha}^*=M_{\beta\alpha}+M_{\alpha\beta},
\end{equation}
or
\begin{equation}
(M^T+M)^*=M+M^T,
\end{equation}
i.e. $M^T+M$ is a real symmetric matrix, and hence it is guaranteed that it has real eigenvectors. Thus the $G$ we obtain in this way will be automatically Hermitian operators.

Alternatively, we do not need to explicitly write $G_n$ in the form of Eq.~(\ref{eq:Gansatz}) and solve the eigenvalues and eigenvectors of $M^T+M$. Instead, we could simply solve the eigenvalue problem of $\mathcal{F}^T+\mathcal{F}$, as stated in the main text, and its eigenvectors will be guaranteed to be Hermitian operators up to an arbitrary overall phase, which can be proved as follows. One could see that $M^T+M$ and $\mathcal{F}^T+\mathcal{F}$ are mapped to each other by 
\begin{equation}
(M^T+M)_{\alpha\beta}=v_{\alpha}^T\cdot (\mathcal{F}^T+\mathcal{F})\cdot v_{\beta},
\end{equation}
where $v_{\alpha}$ is a $N$-site Hermitian basis operator defined previously. Therefore, the eigenvalue problem of $M^T+M$ in Eq.~(\ref{eq:eigM}) is equivalent to
\begin{equation}
U^T(\mathcal{F}^T+\mathcal{F})U\cdot\mathbf{c}=2\tilde{\lambda}_{min}\mathbf{c}
\end{equation}
or
\begin{equation}
(\mathcal{F}^T+\mathcal{F})U\cdot\mathbf{c}=2\lambda_{min}U\cdot\mathbf{c},
\end{equation}
where $\lambda_{min}=2^N\tilde{\lambda}_{min}$ and $U=(v_1,\dots,v_{d^{2N}})$ is a basis transformation in the $d^{2N}$-dimensional complex space. Thus $U\cdot \mathbf{c}$ gives the $G$ in Eq.~(\ref{eq:Gansatz}). Because $\mathbf{c}$ is real and $v_{\alpha}$ is Hermitian, the eigenvectors of $\mathcal{F}^T+\mathcal{F}$ are guaranteed to be Hermitian.

\section{Imposing microscopic symmetries}
Imposing the symmetries of the microscopic Hamiltonian will help to largely reduce the number of variational parameters and thus lower the complexity of the eigenvalue problem.

If the critical one-dimensional system is described by a CFT, the emergent conserved charges will be the zero modes of the Kac-Moody generators, $J_0^{\alpha}$ and $\bar{J}_0^{\alpha}$. For $J_0^{\alpha}+\bar{J}_0^{\alpha}$, it would be time reversal odd and parity even; for $J_0^{\alpha}-\bar{J}_0^{\alpha}$, it would be time reversal even and parity odd. In addition, quantum spin Hamiltonians usually also have spin flip symmetries, and both $J_0^{\alpha}$ and $\bar{J}_0^{\alpha}$ will transform in the same way under the spin flip.

If we write $G$ in the form of Eq.~(\ref{eq:Gansatz}) and solve the eigenvalue problem of $M^T+M$ to optimize $\mathbf{c}$, it will be easy to impose microscopic symmetries. For example, if $d=2$, since $\mathcal{T}\vec{\sigma}\mathcal{T}^{-1}=-\vec{\sigma}$, to impose time-reversal symmetry, we can require the sum in Eq.~(\ref{eq:Gansatz}) to only include $v_{\alpha}$ with odd (even) number of Pauli matrices to make $G$ odd (even) under the time reversal; to impose parity symmetry, we can require the coefficients $c_{\alpha}$ to be $-c_{\beta}$ ($c_{\beta}$) if $v_{\alpha}\xrightarrow{\text{parity}}v_{\beta}$ to make $G$ odd (even) under the parity transformation.

If we use the $\mathcal{F}^T+\mathcal{F}$ formalism, in principle we could restrict the entries in $G$ similarly. But a more convenient way to impose microscopic symmetries when $N$ is not large is to apply projection operators. For example, to make the eigenvector time-reversal even, one could solve the eigenvalue problem of $\mathcal{P}_{+}(\mathcal{F}^T+\mathcal{F})\mathcal{P}_{+}$ instead, where $\mathcal{P}_{+}$ is the projection operator onto the time-reversal even subspace. The operation $\mathcal{P}_{+}\cdot\mathbf{x}$ could be implemented by mapping $\mathbf{x}$ to $\mathbf{x}+\mathcal{T}\mathbf{x}$. In this way, all the time-reversal odd $G$ will move to the null space. When $N$ is not large, we can solve all the eigenvalues from the largest magnitude until we reach the null space and obtain the associated eigenvectors, and by the way we also exclude all the trivial solutions.

\section{Removing trivial solutions}
As mentioned in footnote~\cite{Note2} of the main text, there is an issue to be remembered--the number of trivial solutions is large, i.e. $d^{2(N-1)}$, where $d$ is the dimension of the local Hilbert space, so it would be desired to remove those trivial solutions.

It is easy to remove the trivial solutions in the $M^T+M$ formalism by restricting the operator strings $v_{\alpha}$ involved in Eq.~(\ref{eq:Gansatz}). We first need to exclude the $N$-site identity in Eq.~(\ref{eq:Gansatz}). To further avoid the $p=0$ trivial solutions of the form $X\otimes I-I\otimes X$ (similarly for $p=\pi$), we can fix the coefficient in front of the term $X\otimes I$ and $I\otimes X$ in Eq.~(\ref{eq:Gansatz}) to be the same, without losing any generality, since it simply means that those terms can only appear as $N-1$ site $G=X$ in a symmetric form in the $N$-site $G$.

In the $\mathcal{F}^T+\mathcal{F}$ formalism, as mentioned in the last section, when $N$ is small, one could simply solve all the eigenvalues above the null space to exclude the trivial solutions. More generally, one could get rid of these trivial solutions by lifting those trivial solutions to the top of the spectrum, i.e. $\mathcal{F}\rightarrow \mathcal{F}+\alpha\mathcal{P}_{trivial}$, where $a$ is some positive real number larger than $\lambda_{max}$ of $\mathcal{F}$ and $\mathcal{P}_{trivial}$ is the projection operator onto the trivial solution subspace. To obtain $\mathcal{P}_{trivial}$, we can consider the following problem to find the projection of a given operator $G$ onto the trivial solution subspace, i.e. 
\begin{equation}
    \min_{X}\|G-(X\otimes I-I\otimes X)\|_2,
\end{equation}
where $X\otimes I-I\otimes X$ is the form of the trivial solution at $p=0$. By differentiating with respect to $X$, we get
\begin{equation}
X =\frac{1}{2d}\left(
\begin{diagram}
\draw[rounded corners] (1,2) rectangle (5,1);
\draw (3,1.5) node(X) {$G$};
\draw (3,.75) node {$\cdots$};
\draw (3,2.25) node {$\cdots$};
\draw (1.5,1) -- (1.5,.5); \draw (1.5,2) -- (1.5,2.5);  
\draw (2.5,1) -- (2.5,.5); \draw (2.5,2) -- (2.5,2.5); 
\draw (3.5,1) -- (3.5,.5); \draw (3.5,2) -- (3.5,2.5);
\draw (4.5,2) -- (4.5,2.3); \draw (4.5,1) -- (4.5,.7);
\draw (4.5,2.3) edge[out=90,in=180] (5,2.8);
\draw (4.5,0.7) edge[out=270,in=180] (5,0.2);
\draw (5,2.8) edge[out=0, in=90] (5.5,2.3);
\draw (5,0.2) edge[out=0,in=270] (5.5,0.7);
\draw (5.5,2.3) -- (5.5,.7);
\end{diagram}
-
\begin{diagram}
\draw[rounded corners] (2,2) rectangle (6,1);
\draw (4,1.5) node {$G$};
\draw (3,.75) node {$\cdots$};
\draw (3,2.25) node {$\cdots$};
\draw (5.5,1) -- (5.5,.5); \draw (5.5,2) -- (5.5,2.5);  
\draw (4.5,1) -- (4.5,.5); \draw (4.5,2) -- (4.5,2.5); 
\draw (3.5,1) -- (3.5,.5); \draw (3.5,2) -- (3.5,2.5);
\draw (2.5,2) -- (2.5,2.3); \draw (2.5,1) -- (2.5,.7);
\draw (1.5,2.3) edge[out=90,in=180] (2,2.8);
\draw (1.5,0.7) edge[out=270,in=180] (2,0.2);
\draw (2,2.8) edge[out=0, in=90] (2.5,2.3);
\draw (2,0.2) edge[out=0,in=270] (2.5,0.7);
\draw (1.5,2.3) -- (1.5,.7);
\end{diagram}
+
\begin{diagram}
\draw[rounded corners] (2,2) rectangle (5,1);
\draw (3.5,1.5) node {$X$};
\draw (3,.75) node {$\cdots$};
\draw (3,2.25) node {$\cdots$};
\draw (1.5,2.5) -- (1.5,.5);  
\draw (2.5,1) -- (2.5,.5); \draw (2.5,2) -- (2.5,2.5); 
\draw (3.5,1) -- (3.5,.5); \draw (3.5,2) -- (3.5,2.5);
\draw (4.5,2) -- (4.5,2.3); \draw (4.5,1) -- (4.5,.7);
\draw (4.5,2.3) edge[out=90,in=180] (5,2.8);
\draw (4.5,0.7) edge[out=270,in=180] (5,0.2);
\draw (5,2.8) edge[out=0, in=90] (5.5,2.3);
\draw (5,0.2) edge[out=0,in=270] (5.5,0.7);
\draw (5.5,2.3) -- (5.5,.7);
\end{diagram}
+
\begin{diagram}
\draw[rounded corners] (2,2) rectangle (5,1);
\draw (3.5,1.5) node {$X$};
\draw (3,.75) node {$\cdots$};
\draw (3,2.25) node {$\cdots$};
\draw (5.5,2.5) -- (5.5,.5);  
\draw (4.5,1) -- (4.5,.5); \draw (4.5,2) -- (4.5,2.5); 
\draw (3.5,1) -- (3.5,.5); \draw (3.5,2) -- (3.5,2.5);
\draw (2.5,2) -- (2.5,2.3); \draw (2.5,1) -- (2.5,.7);
\draw (1.5,2.3) edge[out=90,in=180] (2,2.8);
\draw (1.5,0.7) edge[out=270,in=180] (2,0.2);
\draw (2,2.8) edge[out=0, in=90] (2.5,2.3);
\draw (2,0.2) edge[out=0,in=270] (2.5,0.7);
\draw (1.5,2.3) -- (1.5,.7);
\end{diagram}
\right),
\end{equation}
which gives us a linear equation for $X$. If we map an operator $X$ to a state $|X\rangle$, the above equation becomes
\begin{equation}
    A|X\rangle=|b\rangle,
\end{equation}
where
\begin{equation}
A = 
\begin{diagram}
\draw (1.5,2.5) -- (1.5,.5);
\draw (1.8,2.5) -- (1.8,.5);
\draw (2.3,2.5) -- (2.3,.5);
\draw (2.6,2.5) -- (2.6,.5);
\draw (2.85,1.5) node {...};
\draw (3.1,2.5) -- (3.1,.5);
\draw (3.4,2.5) -- (3.4,.5);
\end{diagram}
-\frac{1}{2d}\left(
\begin{diagram}
\draw (3.1,0.5) edge[out=90,in=180] (3.25,1.3);
\draw (3.25,1.3) edge[out=0, in=90] (3.4,0.5);
\draw (1.5,2.5) edge[out=270,in=180] (1.65,1.7);
\draw (1.65,1.7) edge[out=0,in=270] (1.8,2.5);
\draw (1.5,1) -- (1.5,.5);
\draw (1.8,1) -- (1.8,.5);
\draw (2.3,1) -- (2.3,.5);
\draw (2.6,1) -- (2.6,.5);
\draw (2.05,0.75) node {...};
\draw (2.3,2.5) -- (2.3,2);
\draw (2.6,2.5) -- (2.6,2);
\draw (3.1,2.5) -- (3.1,2);
\draw (3.4,2.5) -- (3.4,2);
\draw (2.85,2.25) node {...};
\draw (1.5,1) edge[out=90,in=270] (2.3,2);
\draw (1.8,1) edge[out=90,in=270] (2.6,2);
\draw (2.3,1) edge[out=90,in=270] (3.1,2);
\draw (2.6,1) edge[out=90,in=270] (3.4,2);
\end{diagram}
+
\begin{diagram}
\draw (1.5,0.5) edge[out=90,in=180] (1.65,1.3);
\draw (1.65,1.3) edge[out=0, in=90] (1.8,0.5);
\draw (3.1,2.5) edge[out=270,in=180] (3.25,1.7);
\draw (3.25,1.7) edge[out=0,in=270] (3.4,2.5);
\draw (1.5,2.5) -- (1.5,2);
\draw (1.8,2.5) -- (1.8,2);
\draw (2.3,2.5) -- (2.3,2);
\draw (2.6,2.5) -- (2.6,2);
\draw (2.05,2.25) node {...};
\draw (2.3,1) -- (2.3,.5);
\draw (2.6,1) -- (2.6,.5);
\draw (3.1,1) -- (3.1,.5);
\draw (3.4,1) -- (3.4,.5);
\draw (2.85,0.75) node {...};
\draw (1.5,2) edge[out=270,in=90] (2.3,1);
\draw (1.8,2) edge[out=270,in=90] (2.6,1);
\draw (2.3,2) edge[out=270,in=90] (3.1,1);
\draw (2.6,2) edge[out=270,in=90] (3.4,1);
\end{diagram}
\right)
\end{equation}
and
\begin{equation}
|X\rangle =
\begin{diagram}
\draw[rounded corners] (1,2) rectangle (3.9,1);
\draw (2.45,1.5) node {$X$};
\draw (1.5,2.5) -- (1.5,.5);\draw (1.5,.5) edge[out=270,in=270] (1.8,.5);
\draw (1.8,1) -- (1.8,.5);\draw (1.8,2.5) -- (1.8,2);
\draw (2.3,2.5) -- (2.3,.5);\draw (2.3,.5) edge[out=270,in=270] (2.6,.5);
\draw (2.6,1) -- (2.6,.5);\draw (2.6,2.5) -- (2.6,2);
\draw (2.85,2.25) node {...};
\draw (2.85,0.75) node {...};
\draw (3.1,2.5) -- (3.1,.5);\draw (3.1,.5) edge[out=270,in=270] (3.4,.5);
\draw (3.4,1) -- (3.4,.5);\draw (3.4,2.5) -- (3.4,2);
\end{diagram}
\end{equation}
and
\begin{equation}
|b\rangle =\frac{1}{2d}\left(
\begin{diagram}
\draw[rounded corners] (1,2) rectangle (4.4,1);
\draw (2.7,1.5) node {$G$};
\draw (2.85,.75) node {...};
\draw (2.85,2.25) node {...};
\draw (1.5,2.5) -- (1.5,.5);
\draw (1.8,1) -- (1.8,.5); \draw (1.8,2) -- (1.8,2.5); 
\draw (1.5,.5) edge[out=270,in=270] (1.8,.5);
\draw (2.3,2.5) -- (2.3,.5);
\draw (2.6,1) -- (2.6,.5); \draw (2.6,2) -- (2.6,2.5);
\draw (2.3,.5) edge[out=270,in=270] (2.6,.5);
\draw (3.1,2.5) -- (3.1,.5);
\draw (3.4,1) -- (3.4,.5); \draw (3.4,2) -- (3.4,2.5);
\draw (3.1,.5) edge[out=270,in=270] (3.4,.5);
\draw (3.9,2) -- (3.9,2.3); \draw (3.9,1) -- (3.9,.7);
\draw (3.9,2.3) edge[out=90,in=180] (4.4,2.8);
\draw (3.9,0.7) edge[out=270,in=180] (4.4,0.2);
\draw (4.4,2.8) edge[out=0, in=90] (4.9,2.3);
\draw (4.4,0.2) edge[out=0,in=270] (4.9,0.7);
\draw (4.9,2.3) -- (4.9,.7);
\end{diagram}
-
\begin{diagram}
\draw[rounded corners] (2,2) rectangle (5.4,1);
\draw (3.7,1.5) node {$G$};
\draw (3.55,.75) node {...};
\draw (3.55,2.25) node {...};
\draw (4.9,1) -- (4.9,.5); \draw (4.9,2) -- (4.9,2.5);
\draw (4.6,2.5) -- (4.6,.5);
\draw (4.6,.5) edge[out=270,in=270] (4.9,.5);
\draw (4.1,1) -- (4.1,.5); \draw (4.1,2) -- (4.1,2.5);
\draw (3.8,2.5) -- (3.8,.5);
\draw (3.8,.5) edge[out=270,in=270] (4.1,.5);
\draw (3.3,1) -- (3.3,.5); \draw (3.3,2) -- (3.3,2.5);
\draw (3,2.5) -- (3,.5);
\draw (3,.5) edge[out=270,in=270] (3.3,.5);
\draw (2.5,2) -- (2.5,2.3); \draw (2.5,1) -- (2.5,.7);
\draw (1.5,2.3) edge[out=90,in=180] (2,2.8);
\draw (1.5,0.7) edge[out=270,in=180] (2,0.2);
\draw (2,2.8) edge[out=0, in=90] (2.5,2.3);
\draw (2,0.2) edge[out=0,in=270] (2.5,0.7);
\draw (1.5,2.3) -- (1.5,.7);
\end{diagram}
\right).
\end{equation}
$A$ has a null vector $|I\rangle$ since $A|I\rangle=0$. But $I\otimes I-I\otimes I=0$, so in the solution we can let the coefficient in front of the null vector to be zero without losing any generality. As a result, we get $|X\rangle=\texttt{pinv}(A)|b\rangle$, where $\texttt{pinv}(A)$ is the pseudo-inverse of $A$. Finally, the projection operator onto the trivial solution subspace should be
\begin{equation}
\mathcal{P}_{trivial}=\frac{1}{2d}
\left(
\begin{diagram}
\draw[rounded corners] (1.2,2) rectangle (3.7,1);
\draw (1.5,1) -- (1.5,.5);\draw (1.5,2.5) -- (1.5,2);
\draw (1.8,1) -- (1.8,.5);\draw (1.8,2.5) -- (1.8,2);
\draw (2.3,1) -- (2.3,.5);\draw (2.3,2.5) -- (2.3,2);
\draw (2.6,1) -- (2.6,.5);\draw (2.6,2.5) -- (2.6,2);
\draw (2.45,1.5) node {$\texttt{pinv}(A)$};
\draw (2.85,2.25) node {...};
\draw (2.85,0.75) node {...};
\draw (3.1,1) -- (3.1,.5);\draw (3.1,2.5) -- (3.1,2);
\draw (3.4,1) -- (3.4,.5);\draw (3.4,2.5) -- (3.4,2);
\draw (3.9,0.5) edge[out=90,in=180] (4.05,1.3);
\draw (4.05,1.3) edge[out=0, in=90] (4.2,0.5);
\draw (3.9,2.5) edge[out=270,in=180] (4.05,1.7);
\draw (4.05,1.7) edge[out=0,in=270] (4.2,2.5);
\end{diagram}
+
\begin{diagram}
\draw (1.5,0.5) edge[out=90,in=180] (1.65,1.3);
\draw (1.65,1.3) edge[out=0, in=90] (1.8,0.5);
\draw (1.5,2.5) edge[out=270,in=180] (1.65,1.7);
\draw (1.65,1.7) edge[out=0,in=270] (1.8,2.5);
\draw[rounded corners] (2,2) rectangle (4.5,1);
\draw (2.3,1) -- (2.3,.5);\draw (2.3,2.5) -- (2.3,2);
\draw (2.6,1) -- (2.6,.5);\draw (2.6,2.5) -- (2.6,2);
\draw (3.1,1) -- (3.1,.5);\draw (3.1,2.5) -- (3.1,2);
\draw (3.4,1) -- (3.4,.5);\draw (3.4,2.5) -- (3.4,2);
\draw (3.25,1.5) node {$\texttt{pinv}(A)$};
\draw (3.65,2.25) node {...};
\draw (3.65,0.75) node {...};
\draw (3.9,1) -- (3.9,.5);\draw (3.9,2.5) -- (3.9,2);
\draw (4.2,1) -- (4.2,.5);\draw (4.2,2.5) -- (4.2,2);
\end{diagram}
-
\begin{diagram}
\draw (3.9,0.5) edge[out=90,in=180] (4.05,1.3);
\draw (4.05,1.3) edge[out=0, in=90] (4.2,0.5);
\draw (1.5,2.5) edge[out=270,in=180] (1.65,1.7);
\draw (1.65,1.7) edge[out=0,in=270] (1.8,2.5);
\draw[rounded corners] (1.9,2) rectangle (3.8,1);
\draw (2.85,1.5) node {$\texttt{pinv}(A)$};
\draw (1.5,1) -- (1.5,.5);
\draw (1.8,1) -- (1.8,.5);
\draw (2.3,1) -- (2.3,.5);
\draw (2.6,1) -- (2.6,.5);
\draw (2.85,0.75) node {...};
\draw (3.1,1) -- (3.1,.5);
\draw (3.4,1) -- (3.4,.5);
\draw (2.3,2.5) -- (2.3,2);
\draw (2.6,2.5) -- (2.6,2);
\draw (3.1,2.5) -- (3.1,2);
\draw (3.4,2.5) -- (3.4,2);
\draw (3.65,2.25) node {...};
\draw (3.9,2.5) -- (3.9,2);
\draw (4.2,2.5) -- (4.2,2);
\draw (1.5,1) edge[out=90,in=220] (1.9,1.6);
\draw (1.8,1) edge[out=90,in=220] (1.9,1.3);
\draw (3.8,1.6) edge[out=40,in=270] (3.9,2);
\draw (3.8,1.3) edge[out=40,in=270] (4.2,2);
\end{diagram}
-
\begin{diagram}
\draw (1.5,0.5) edge[out=90,in=180] (1.65,1.3);
\draw (1.65,1.3) edge[out=0, in=90] (1.8,0.5);
\draw (3.9,2.5) edge[out=270,in=180] (4.05,1.7);
\draw (4.05,1.7) edge[out=0,in=270] (4.2,2.5);
\draw[rounded corners] (1.9,2) rectangle (3.8,1);
\draw (2.85,1.5) node {$\texttt{pinv}(A)$};
\draw (1.5,2.5) -- (1.5,2);
\draw (1.8,2.5) -- (1.8,2);
\draw (2.3,2.5) -- (2.3,2);
\draw (2.6,2.5) -- (2.6,2);
\draw (2.85,2.25) node {...};
\draw (3.1,2.5) -- (3.1,2);
\draw (3.4,2.5) -- (3.4,2);
\draw (2.3,1) -- (2.3,.5);
\draw (2.6,1) -- (2.6,.5);
\draw (3.1,1) -- (3.1,.5);
\draw (3.4,1) -- (3.4,.5);
\draw (3.65,0.75) node {...};
\draw (3.9,1) -- (3.9,.5);
\draw (4.2,1) -- (4.2,.5);
\draw (1.5,2) edge[out=270,in=140] (1.9,1.3);
\draw (1.8,2) edge[out=270,in=140] (1.9,1.6);
\draw (3.8,1.3) edge[out=320,in=90] (3.9,1);
\draw (3.8,1.6) edge[out=320,in=90] (4.2,1);
\end{diagram}
\right).
\end{equation}
Then we can perform $\mathcal{F}\rightarrow -(\mathcal{F}+\alpha\mathcal{P}_{trivial})$ to shift the trivial solution subspace to the top and reverse the spectrum so that the problem transforms to solving for the largest eigenvalues. $\mathcal{P}_{trivial}$ is model independent, so for each $N$ we only need to solve for $\texttt{pinv}(A)$ once.

\section{Finite-system DMRG to optimize $G$}
In the $\mathcal{F}^T+\mathcal{F}$ formalism, if we treat $G$ as a single big tensor, the complexity of the optimization problem will grow exponentially as $N$ increases. Alternatively, we can write $G$ as a finite matrix product operator (MPO) and use the density matrix renormalization group (DMRG)~\cite{PhysRevLett.69.2863,PhysRevB.48.10345} to optimize it. Note that if we transform to the $M^T+M$ formalism, i.e. using the operator string $v_{\alpha}$ basis, the coefficient $c_{\alpha}$ can also be expressed as an MPO, and the advantage of using this basis is that the eigenvalue problem can be solved within real numbers. 

Now let's take $N=5$ as an example. A 5-site finite MPO is illustrated as
\begin{equation}
G =
\begin{diagram}
\draw[rounded corners] (1,2) rectangle (2,1);
\draw (1.5,1.5) node (X) {$W_1$};
\draw (2,1.5) -- (3,1.5); 
\draw[rounded corners] (3,2) rectangle (4,1);
\draw (3.5,1.5) node {$W_2$};
\draw (4,1.5) -- (5,1.5);
\draw[rounded corners] (5,2) rectangle (6,1);
\draw (5.5,1.5) node {$W_3$};
\draw (6,1.5) -- (7,1.5); 
\draw[rounded corners] (7,2) rectangle (8,1);
\draw (7.5,1.5) node {$W_4$};
\draw (8,1.5) -- (9,1.5); 
\draw[rounded corners] (9,2) rectangle (10,1);
\draw (9.5,1.5) node {$W_5$};
\draw (1.5,1) -- (1.5,.5);\draw (1.5,2.5) -- (1.5,2);
\draw (3.5,1) -- (3.5,.5);\draw (3.5,2.5) -- (3.5,2);
\draw (5.5,1) -- (5.5,.5);\draw (5.5,2.5) -- (5.5,2);
\draw (7.5,1) -- (7.5,.5);\draw (7.5,2.5) -- (7.5,2);
\draw (9.5,1) -- (9.5,.5);\draw (9.5,2.5) -- (9.5,2); 
\end{diagram}.
\end{equation}
Different from treating $G$ as a single big tensor, at each iteration we assume all but one site tensor $W_i$ constant and differentiate the cost function $f$ with respect to $W_i$ only. If we let $G$ in its canonical form, just like an MPS, we will get an eigenvalue problem for $W_i$ at each iteration. We solve for the lowest eigenvalue at each iteration, and after several DMRG sweeps through the 5 sites we get a $G$ corresponding to $\lambda_{min}$. Then we can construct a projection operator $\mathcal{P}=|G\rangle\langle G|$. Modifying the cost function to be $\mathcal{F}+\beta \mathcal{P}$ to lift this solution to the top of the spectrum, we perform the DMRG again to solve for the next solution.

Removing the large number of trivial solutions is essential for doing DMRG efficiently. In principle, we could do the same thing as in the last section. However, performing efficient DMRG requires us to have $\texttt{pinv}(A)$ either in the form of an MPO or decomposition of some local operators, while we currently do not have a good approach to efficiently calculate $\texttt{pinv}(A)$ when $N$ becomes larger.

\section{\label{app:subsec}The infinite uniform MPO formalism}
A more natural representation for $O$, which is able to hold long-range interacting terms in the summation, would be the finite state automaton~\cite{PhysRevA.78.012356,PhysRevB.78.035116}, which can be translated to a infinite uniform MPO. This formalism also helps to reduce the complexity of the problem by solving for only one MPO site tensor and restricting the form of $G$ to certain combination of Pauli strings. Here, we explain how to optimize a MPO for $O$ of bond dimension $\chi_W = 2$, and show that the optimization for this MPO gives the same eigenvalue problem as in the main text. However, generalization from $\chi_W = 2$ to $\chi_W > 2$ is non-trivial and we leave it as an open question.

The MPO representation of $O$ is
\begin{equation} 
O =  \dots
\begin{diagram}
\draw (0.5,1.5) -- (1,1.5); 
\draw[rounded corners] (1,2) rectangle (2,1);
\draw (1.5,1.5) node (X) {$W$};
\draw (2,1.5) -- (3,1.5); 
\draw[rounded corners] (3,2) rectangle (4,1);
\draw (3.5,1.5) node {$W$};
\draw (4,1.5) -- (5,1.5);
\draw[rounded corners] (5,2) rectangle (6,1);
\draw (5.5,1.5) node {$W$};
\draw (6,1.5) -- (7,1.5); 
\draw[rounded corners] (7,2) rectangle (8,1);
\draw (7.5,1.5) node {$W$};
\draw (8,1.5) -- (9,1.5); 
\draw[rounded corners] (9,2) rectangle (10,1);
\draw (9.5,1.5) node {$W$};
\draw (10,1.5) -- (10.5,1.5); 
\draw (1.5,1) -- (1.5,.5);\draw (1.5,2.5) -- (1.5,2);
\draw (3.5,1) -- (3.5,.5);\draw (3.5,2.5) -- (3.5,2);
\draw (5.5,1) -- (5.5,.5);\draw (5.5,2.5) -- (5.5,2);
\draw (7.5,1) -- (7.5,.5);\draw (7.5,2.5) -- (7.5,2);
\draw (9.5,1) -- (9.5,.5);\draw (9.5,2.5) -- (9.5,2); 
\end{diagram} \dots,
\end{equation}
where $W$ is an operator-valued matrix given by
\begin{equation}
W = \left[
\begin{matrix}
\mathds{1}& G\\
0&\mathds{1}
\end{matrix}
\right].
\end{equation}
Then
\begin{equation}
\langle\psi |\frac{\partial O^{\dag}}{\partial G^{\dag}}O|\psi\rangle=
\begin{diagram}
\draw[rounded corners] (0.7,4.9) rectangle (2.3,0);
\draw (1.5,2.45) node (X)  {$L^{[\bar{W}W]}$};
\draw (3.3,0.5) node  {$\bar{A}_C$};
\draw (3.3,3.1) node  {$W$};
\draw (3.3,4.4) node  {$A_C$};
\draw[rounded corners] (2.8,1) rectangle (3.8,0);
\draw[rounded corners] (2.8,3.6) rectangle (3.8,2.6);
\draw[rounded corners] (2.8,4.9) rectangle (3.8,3.9);
\draw[rounded corners] (5.9,4.9) rectangle (4.3,0);
\draw (5.1,2.45) node {$R^{[\bar{W}W]}$};
\draw (2.3,.5) -- (2.8,.5);\draw (3.8,.5) -- (4.3,.5);
\draw (2.3,1.8) -- (2.8,1.8);\draw (3.8,1.8) -- (4.3,1.8);
\draw (2.3,4.4) -- (2.8,4.4);\draw (3.8,4.4) -- (4.3,4.4);
\draw (3.3,1) -- (3.3,1.3);\draw (3.3,2.3) -- (3.3,2.6);\draw (3.3,3.6) -- (3.3,3.9);
\draw (2.3,3.1) -- (2.8,3.1);\draw (3.8,3.1) -- (4.3,3.1);
\draw (2.9,1.8) node {1};\draw (3.7,1.8) node {2};
\end{diagram}
=D^{12,11}W_{11}+D^{12,12}W_{12}+D^{12,21}W_{21}+D^{12,22}W_{22}=D^{12,11}\mathds{1}+D^{12,12}G+D^{12,22}\mathds{1}.
\end{equation}
Therefore we only need to calculate the fixed points $(L^{[\bar{W}{W}]}_{1,1}|$, $|R^{[\bar{W}{W}]}_{2,1})$, $(L^{[\bar{W}{W}]}_{1,2}|$, $|R^{[\bar{W}{W}]}_{2,2})$, which are defined as
\begin{equation}
(L^{[\bar{W}{W}]}_{a,b}|=\sum_{(a',b')\leq(a,b)}(L^{[\bar{W}{W}]}_{a',b'}|(T^{[\bar{W}W]}_L)_{a',b';a,b},~~|R^{[\bar{W}{W}]}_{a,b})=\sum_{(a',b')\geq (a,b)}(T^{[\bar{W}W]}_R)_{a,b;a',b'}|R^{[\bar{W}{W}]}_{a',b'}),
\end{equation}
where
\begin{equation}
(T^{[\bar{W}W]}_{L/R})_{a',b';a,b}=\sum_{s,s',s''} \bar{A}_{L/R}^{s''}\otimes\bar{W}_{a',a}^{s',s''}\otimes W_{b',b}^{s,s'}\otimes A_{L/R}^s.
\end{equation}
Notice that $\bar{W}\otimes W$ is still upper-triangular and there are two additional identities in the diagonal elements, i.e.
\begin{equation}
\bar{W}\otimes W = \left[
\begin{matrix}
\mathds{1}& G & G^{\dag} & \mathds{1}\\
0 & \mathds{1} & 0 & G^{\dag}\\
0 & 0 & \mathds{1} & G \\
0 & 0 & 0 & \mathds{1}
\end{matrix}
\right].   
\end{equation}
From the definition, substituting the form of $W$, we get $(L^{[\bar{W}{W}]}_{1,1}|=(L^{[\bar{W}{W}]}_{1,1}|T_L$, where $T_L$ is the transfer operator of the MPS $|\psi [A_L]\rangle$, so we have $(L^{[\bar{W}{W}]}_{1,1}|=(1|$, where $(1|$ is the left fixed point of $T_L$. We can then get $(L^{[\bar{W}{W}]}_{1,2}|=(L^{[\bar{W}{W}]}_{1,2}|T_L+(Y_{1,2}|$, where 
\begin{equation}
(Y_{1,2}|=
\begin{diagram}
\draw (1,1) edge[out=180,in=180] (1,4);
\draw[rounded corners] (1,3.5) rectangle (2,4.5);
\draw[rounded corners] (1,0.5) rectangle (2,1.5);
\draw[rounded corners] (1,2) rectangle (2,3);
\draw (1.5,2.5) node (X) {$G$};
\draw (1.5,1) node {$\bar{A}_L$};
\draw (1.5,4) node {$A_L$};
\draw (1.5,1.5) -- (1.5,2);
\draw (1.5,3) -- (1.5,3.5);
\draw (2,4) -- (3,4); \draw (2,1) -- (3,1); 
\end{diagram}.
\end{equation}
To remove the divergence from $\langle O\rangle$, one needs instead to solve the linear equation
\begin{equation}
(L^{[\bar{W}{W}]}_{1,2}|(1-T_L+|R)(1|)=(Y_{1,2}|-(Y_{1,2}|R)(1|
\end{equation}
to get $(L^{[\bar{W}{W}]}_{1,2}|$. Similarly we can get $|R^{[\bar{W}{W}]}_{2,2})=|1)$ and 
\begin{equation}
(1-T_R+|1)(L|)|R^{[\bar{W}{W}]}_{2,1})=|Y_{2,1})-|1)(L|Y_{2,1}),
\end{equation}
where
\begin{equation}
|Y_{2,1})=
\begin{diagram}
\draw (2,1) edge[out=0,in=0] (2,4);
\draw[rounded corners] (1,3.5) rectangle (2,4.5);
\draw[rounded corners] (1,0.5) rectangle (2,1.5);
\draw[rounded corners] (1,2) rectangle (2,3);
\draw (1.5,2.5) node (X) {$G$};
\draw (1.5,1) node {$\bar{A}_R$};
\draw (1.5,4) node {$A_R$};
\draw (1.5,1.5) -- (1.5,2);
\draw (1.5,3) -- (1.5,3.5);
\draw (0,4) -- (1,4); \draw (0,1) -- (1,1); 
\end{diagram}.
\end{equation}
Therefore $D^{12,11}$ and $D^{12,22}$ are proportional to $G$ and $D^{12,12}$ has no $G$ dependence. It is easy to see that $D^{12,11}\mathds{1}+D^{12,12}G+D^{12,22}\mathds{1}$ is equivalent to $\mathcal{F}\cdot\mathbf{g}$ in the main text. If we require the normalization $\mathrm{Tr}[G^{\dag}G]=1$, then we get the same eigenvalue problem as in the main text. So the $\chi_W = 2$ MPO formalism is equivalent to representing $O$ as a summation of local terms in the case that $G$ is one-site.

However, this equivalence could not be generalized to $\chi_W>2$. For example, the most generic form for $\chi_W=3$ is
\begin{equation}
W = \left[
\begin{matrix}
\mathds{1}& A & B\\
0 & D & C\\
0 & 0 &\mathds{1}
\end{matrix}
\right].
\end{equation}
If we want there to be some exponentially decaying interacting term, then $D=\kappa\mathds{1}$ with $\kappa<1$. There are several issues to optimize $W$. First, considering the simplest case with $D=0$, the $O$ generated by $W$ would be $\sum_n(A_nC_{n+1}+B_n)$, and $\langle\psi |O^{\dag}O|\psi\rangle$ contain terms linear in $A$, $B$, and $C$, so taking its derivative with respect to $A$, $B$, or $C$ would not give us an eigenvalue problem, but one might be able to use the gradient descent method. The second question would be how to choose a proper normalization condition for $W$, especially when there is an exponentially decaying term. Simply requiring $A$, $B$, and $C$ each to be individually normalized does not make sense. Probably the canonical form~\cite{PhysRevB.102.035147} for such Hamiltonian-like MPO might help.

\section{Method based on the fundamental theorem of MPS}
In footnote~\cite{Note3} of the main text, we mentioned that solving the symmetry of the ground state of a gapped system is much easier, and another method based on the MPS fundamental theorem also works. In this section, we will discuss this alternative method for gapped systems and why it does not work well for critical systems.

If an MPS
\begin{equation} 
\ket{\Psi(A)} =  \dots
\begin{diagram}
\draw (0.5,1.5) -- (1,1.5); 
\draw[rounded corners] (1,2) rectangle (2,1);
\draw (1.5,1.5) node (X) {$A$};
\draw (2,1.5) -- (3,1.5); 
\draw[rounded corners] (3,2) rectangle (4,1);
\draw (3.5,1.5) node {$A$};
\draw (4,1.5) -- (5,1.5);
\draw[rounded corners] (5,2) rectangle (6,1);
\draw (5.5,1.5) node {$A$};
\draw (6,1.5) -- (7,1.5); 
\draw[rounded corners] (7,2) rectangle (8,1);
\draw (7.5,1.5) node {$A$};
\draw (8,1.5) -- (9,1.5); 
\draw[rounded corners] (9,2) rectangle (10,1);
\draw (9.5,1.5) node {$A$};
\draw (10,1.5) -- (10.5,1.5);
\draw (1.5,1) -- (1.5,.5); \draw (3.5,1) -- (3.5,.5); \draw (5.5,1) -- (5.5,.5);
\draw (7.5,1) -- (7.5,.5); \draw (9.5,1) -- (9.5,.5); 
\end{diagram} \dots.
\end{equation}
is symmetric under a global on-site symmetry operation, the MPS tensor itself transforms under this operation~\cite{10.21468/SciPostPhysLectNotes.7}:
\begin{equation}
 U_g^{\otimes N} \ket{\Psi(A)} \propto \ket{\Psi(A)} \qquad \rightarrow \quad 
\begin{diagram} 
 \draw (.5,-1) circle (.5); \draw (.5,-1) node {$U_g$};
 \draw[rounded corners] (0,1) rectangle (1,0);
 \draw (.5,.5) node (X) {$A$};
 \draw (-.5,.5) -- (0,.5); \draw (1,.5) -- (1.5,.5); 
 \draw (.5,0) -- (.5,-.5); \draw (.5,-1.5) -- (.5,-2); 
 \end{diagram}
 =  
 \e^{i\phi_g} \; \begin{diagram} 
 \draw (-1.5,.5) circle (.5); \draw (-1.5,.5) node {$V_g$};
 \draw (2.5,.5) circle (.5); \draw (2.5,.5) node {$V_g\dag$};
 \draw[rounded corners] (0,1) rectangle (1,0);
 \draw (.5,.5) node (X) {$A$};
 \draw (-1,.5) -- (0,.5); \draw (1,.5) -- (2,.5); 
 \draw (-2.5,.5) -- (-2,.5); \draw (3,.5) -- (3.5,.5);
 \draw (.5,-.5) -- (.5,0);
 \end{diagram} \;.
\end{equation}
If the symmetry transformation is continuous, we can write $U_g=e^{i\epsilon G}$. After absorbing the phase factor $e^{i\phi_g}$ into the definition of $G$, the infinitesimal version becomes
\begin{equation}
\begin{diagram} 
 \draw (.5,-1) circle (.5); \draw (.5,-1) node {\scalebox{0.6}{$1+i\epsilon G$}};
 \draw[rounded corners] (0,1) rectangle (1,0);
 \draw (.5,.5) node (X) {$A$};
 \draw (-.5,.5) -- (0,.5); \draw (1,.5) -- (1.5,.5); 
 \draw (.5,0) -- (.5,-.5); \draw (.5,-1.5) -- (.5,-2); 
 \end{diagram}
 =  
\begin{diagram} 
 \draw (-1.5,.5) circle (.5); \draw (-1.5,.5) node {\scalebox{0.6}{$1+i\epsilon K$}};
 \draw (2.5,.5) circle (.5); \draw (2.5,.5) node {\scalebox{0.6}{$1-i\epsilon K$}};
 \draw[rounded corners] (0,1) rectangle (1,0);
 \draw (.5,.5) node (X) {$A$};
 \draw (-1,.5) -- (0,.5); \draw (1,.5) -- (2,.5); 
 \draw (-2.5,.5) -- (-2,.5); \draw (3,.5) -- (3.5,.5);
 \draw (.5,-.5) -- (.5,0); 
 \end{diagram}
\end{equation}
or
\begin{equation} 
\begin{diagram} 
 \draw (.5,-1) circle (.5); \draw (.5,-1) node {$G$};
 \draw[rounded corners] (0,1) rectangle (1,0);
 \draw (.5,.5) node (X) {$A$};
 \draw (-.5,.5) -- (0,.5); \draw (1,.5) -- (1.5,.5); 
 \draw (.5,0) -- (.5,-.5);  \draw (.5,-1.5) -- (.5,-2);
 \end{diagram}
 =
 \begin{diagram} 
 \draw (-1.5,.5) circle (.5); \draw (-1.5,.5) node {$K$};
 \draw[rounded corners] (0,1) rectangle (1,0);
 \draw (.5,.5) node (X) {$A$};
 \draw (-1,.5) -- (0,.5); \draw (1,.5) -- (1.5,.5); 
 \draw (-2.5,.5) -- (-2,.5); 
 \draw (.5,-.5) -- (.5,0);
 \end{diagram}
 -
 \begin{diagram} 
 \draw (2.5,.5) circle (.5); \draw (2.5,.5) node {$K$};
 \draw[rounded corners] (0,1) rectangle (1,0);
 \draw (.5,.5) node (X) {$A$};
 \draw (-.5,.5) -- (0,.5);\draw (1,.5) -- (2,.5); 
 \draw (3,.5) -- (3.5,.5);
 \draw (.5,-.5) -- (.5,0);
 \end{diagram}\;.
\end{equation}
After vectorizing $G\mapsto \mathbf{g}$ and $K\mapsto \mathbf{k}$, the above equation becomes
\begin{equation}\label{eqn:solvegk}
\left[\begin{matrix}
P & Q-R
\end{matrix}\right]
\left[\begin{matrix}
\mathbf{g}\\
\mathbf{k}
\end{matrix}\right]=0,
\end{equation}
where $P_{i\alpha\beta,jl}=\delta_{ij}A^l_{\alpha\beta}$, $Q_{i\alpha\beta,\gamma\rho}=\delta_{\rho\beta}A^i_{\alpha\gamma}$, and $R_{i\alpha\beta,\gamma\rho}=\delta_{\alpha\gamma}A^i_{\rho\beta}$.
The equation can be transformed into a least square problem
\begin{equation}
\min_{\mathbf{x}}\|M\cdot\mathbf{x}\|
\end{equation}
with constraint $\|\mathbf{x}\|_2=1$, where $M = [P~Q-R]$ and $\mathbf{x}=[\mathbf{g}~\mathbf{k}]^T$. This problem can be solved by either quadratic programming or singular value decomposition (SVD) $M=USV^{\dag}$, where $S=\mathrm{diag}(\lambda_1,\cdots,\lambda_{d^2+D^2})$ and $V=(\mathbf{v}_1,\cdots,\mathbf{v}_{d^2+D^2})$.

While it works for gapped systems, this way to solve for the symmetries is numerically unstable for the critical systems. In the benchmarks for the 1D isotropic $XY$ model, it was observed that the singular vector $\mathbf{v}_i$ whose first $d^2$ components best approximate the symmetry generator $S_z$ does not always correspond to the second lowest singular value $\lambda_{d^2+D^2-1}$ (the trivial solution is the lowest one). In addition, the SVD only requires the orthogonality
\begin{equation}
\left[\begin{matrix}
\mathbf{g} & \mathbf{k}
\end{matrix}\right]^{\dag}
\left[\begin{matrix}
\mathbf{g}'\\
\mathbf{k}'
\end{matrix}\right]=0
\end{equation}
but not necessarily $\mathbf{g}^{\dag}\mathbf{g}'=0$, so there are many $\mathbf{v}_j$'s that have nonzero overlap with $(S_z,K)$. Also, we didn't see a clear scaling of the singular values with the correlation length.

Furthermore, $M$ is not gauge invariant and so are its singular values. If we work in the mixed gauge, the infinitesimal version of the fundamental theorem becomes
\begin{equation} 
\begin{diagram} 
 \draw (.5,-1) circle (.5); \draw (.5,-1) node {$G$};
 \draw[rounded corners] (0,1) rectangle (1,0);
 \draw (.5,.5) node (X) {$A_C$};
 \draw (-.5,.5) -- (0,.5); \draw (1,.5) -- (1.5,.5); 
 \draw (.5,0) -- (.5,-.5);  \draw (.5,-1.5) -- (.5,-2);
 \end{diagram}
 =
 \begin{diagram} 
 \draw (-1.5,.5) circle (.5); \draw (-1.5,.5) node {$K$};
 \draw[rounded corners] (0,1) rectangle (1,0);
 \draw (.5,.5) node (X) {$A_R$};
 \draw (-1,.5) -- (0,.5); \draw (1,.5) -- (1.5,.5); 
 \draw (-2.5,.5) -- (-2,.5); 
 \draw (.5,-.5) -- (.5,0);
 \end{diagram}
 -
 \begin{diagram} 
 \draw (2.5,.5) circle (.5); \draw (2.5,.5) node {$K$};
 \draw[rounded corners] (0,1) rectangle (1,0);
 \draw (.5,.5) node (X) {$A_L$};
 \draw (-.5,.5) -- (0,.5);\draw (1,.5) -- (2,.5); 
 \draw (3,.5) -- (3.5,.5);
 \draw (.5,-.5) -- (.5,0);
 \end{diagram}\;
\end{equation}
with $A_C=LAR$, $A_L=LAL^{-1}$, and $A_R=R^{-1}AR$. Notice that the trivial solution now becomes $(G,K)=(0,C)$, where $C=LR$. In the benchmarks for the 1D isotropic $XY$ model, we could see a clear scaling of the singular values with the correlation length $\xi$. However, the problems are only partially solved by utilizing the mixed gauge. $\mathbf{g}$ in the singular vectors is still not orthogonal to each other. As a result, we observed that the number of singular values that scales to zero with increasing correlation length is much larger than the number of symmetries the system actually has, though there is a gap between the lowest ones and the higher ones. The $D^2$ useless degrees of freedom for $K$ also increase the complexity of the problem and make the solution quickly become unreachable as $N$ and $D$ increases. In addition, the singular values are not very physical, since they are still gauge dependent. In conclusion, for critical systems one should resort to the variational method in the main text.

Note that the on-site symmetry $\bigotimes_n U_g=\exp{(-i\epsilon\sum_nG_n)}$ can be generalized to a continuous symmetry $U=\exp{(-i\epsilon O)}$ generated by a summation of $N$-site local Hermitian operators $O=\sum_nG_{n,\cdots,n+N-1}$, which takes a similar form to a local Hamiltonian. If the state is invariant under the transformation, i.e. $\exp{(-i\epsilon O)}|\psi\rangle=|\psi\rangle$ (the phase factor has been absorbed into the definition of $O$), it implies equivalently $O|\psi\rangle=\sum_nG_{n,\cdots,n+N-1}|\psi\rangle=0$, or
\begin{equation}
\begin{diagram}
\draw (1.5,1.5) node (X) {$\phantom{X}$};
\draw[rounded corners] (1,2) rectangle (2,1);
\draw[rounded corners] (3,2) rectangle (4,1);
\draw[rounded corners] (5,2) rectangle (6,1);
\draw[rounded corners] (1,-.5) rectangle (6,.5);
\draw (1.5,1.5) node {$A_L$};\draw (3.5,1.5) node {$A_C$};\draw (5.5,1.5) node {$A_R$};
\draw (3.5,0) node {$G$};
\draw (1.5,.5) -- (1.5,1); \draw (3.5, .5) -- (3.5,1); \draw (5.5,.5) -- (5.5,1);
\draw (1.5,-.5) -- (1.5,-1); \draw (3.5, -.5) -- (3.5,-1); \draw (5.5,-.5) -- (5.5,-1);
\draw (.5,1.5) -- (1,1.5); \draw (6,1.5) -- (6.5,1.5);
\draw (2,1.5) -- (2.2,1.5); \draw (2.5,1.5) node {$\cdots$}; \draw (2.8,1.5) -- (3,1.5);
\draw (4,1.5) -- (4.2,1.5); \draw (4.5,1.5) node {$\cdots$}; \draw (4.8,1.5) -- (5,1.5);
\draw (2.5,.75) node {$\cdots$}; \draw (4.5,.75) node {$\cdots$};
\draw (2.5,-.75) node {$\cdots$}; \draw (4.5,-.75) node {$\cdots$};
\end{diagram}
=
\begin{diagram}
\draw (1.5,1.5) node (X) {$\phantom{X}$};
\draw[rounded corners] (1,2) rectangle (4,1);
\draw[rounded corners] (5,2) rectangle (6,1);
\draw (2.5,1.5) node {$K$}; \draw (5.5,1.5) node {$A_R$};
\draw (2.5,.75) node {$\cdots$}; 
\draw (1.5,.5) -- (1.5,1); \draw (3.5, .5) -- (3.5,1); \draw (5.5,.5) -- (5.5,1);
\draw (.5,1.5) -- (1,1.5);  \draw (4,1.5) -- (5,1.5); \draw (6,1.5) -- (6.5,1.5);
\end{diagram}
-
\begin{diagram}
\draw[rounded corners] (1,2) rectangle (2,1);
\draw[rounded corners] (3,2) rectangle (6,1);
\draw (1.5,1.5) node {$A_L$}; \draw (4.5,1.5) node {$K$};
\draw (4.5,.75) node {$\cdots$};
\draw (1.5,.5) -- (1.5,1); \draw (3.5, .5) -- (3.5,1); \draw (5.5,.5) -- (5.5,1);
\draw (.5,1.5) -- (1,1.5);  \draw (2,1.5) -- (3,1.5); \draw (6,1.5) -- (6.5,1.5);
\end{diagram}\;
\end{equation}
for injective MPS, where $G$ is a $N$-site operator and $K$ is a $N-1$-site tensor. Now there are $d^{2(N-1)}$ trivial solutions, which take the form $G=X\otimes I-I\otimes X$, where $X$ is a $N-1$-site operator and $I$ is the one-site identity operator.

\section{Conserved quantities in the spin-1/2 isotropic quantum $XY$ chain}
In this section, we review the master symmetry~\cite{https://doi.org/10.1002/sapm1985733221} technique to obtain the conserved quantities in the spin-1/2 isotropic quantum $XY$ chain.

The Hamiltonian of the spin-1/2 isotropic quantum $XY$ chain can be written as 
\begin{equation}
H = \sum_ih_i
\end{equation}
where $h_i=X_iX_{i+1}+Y_iY_{i+1}$. Obviously, $H$ has a $\mathrm{U}(1)$ symmetry and thus $[H,Q_0]=0$, where $Q_0=\sum_iZ_i$. Each term in the conserved quantity $Q_1=\sum_i(X_iY_{i+1}-Y_iX_{i+1})$ can be obtained by Eq. (46) in Ref.~\cite{PhysRevB.106.115111}, i.e. $2\mathrm{i}(X_iY_{i+1}-Y_iX_{i+1})=[h_i,Z_i]$. If we denote $H_0=H$, each term in level-$n$ conserved quantities $H_n=\sum_ih_{n,i}$ and $Q_n=\sum_iq_{n,i}$ can be constructed recursively up to a constant by
\begin{subequations}
\begin{eqnarray}
h_{n+1,i} \propto [h_i,H_n],~~~~\mathrm{if}~~n\geq 0;\\
q_{n+1,i} \propto [h_i,Q_n],~~~~\mathrm{if}~~n\geq 1.
\end{eqnarray}
\end{subequations}
For example, we have $H_1=\sum_i(X_iZ_{i+1}Y_{i+2}-Y_iZ_{i+1}X_{i+2})$, and $Q_2=\sum_i(X_iZ_{i+1}X_{i+2}+Y_iZ_{i+1}Y_{i+2})$. One can verify that
\begin{equation}
[H_n,H_m] = [H_n,Q_m] = [Q_n,Q_m] = 0.    
\end{equation}
For $p=\pi$ we could get in a similar way the conserved quantities that take a staggered pattern in the summation, $K=\sum_i(-1)^ik_i$.

\section{Explanation of the critical exponents of the conserved quantities in the spin-1/2 isotropic quantum $XY$ chain}
For a correlation function $C(r,\xi)=\langle\psi |G^{\dag}_{n+r}G_n|\psi\rangle$ near the critical point, it takes the form~\cite{PhysRevX.8.041033}
\begin{equation}
    C(r,\xi)=r^{-\eta}g(r/\xi),
\end{equation}
where $\eta=2\Delta$ is double the scaling dimension of the operator $G$. Since the effect of finite bond dimension is like a relevant perturbation which deforms the MPS away from the exact critical point, the correlation function calculated with MPS should also take this form. In the continuuum limit, the static structure factor $S$ at momentum $p=0$ is given by
\begin{equation}
S(\xi)=2\int_{0}^{\infty}\mathrm{d}r r^{-\eta}g(r/\xi),
\end{equation}
which should be invariant under a change of integration variable $x=\lambda r$, i.e.
\begin{equation}
S(\xi)=2\lambda^{\eta-1}\int_{0}^{\infty}\mathrm{d}x x^{-\eta}g(x/(\lambda\xi))=\lambda^{\eta-1}S(\lambda\xi).  
\end{equation}
Therefore, we have $S(\lambda\xi)=\lambda^{-(\eta-1)}S(\xi)$, which yields
\begin{equation}
S(\xi)\sim\xi^{-(\eta-1)},
\end{equation}
i.e. the critical exponent for $S(\xi)$ is given by the critical exponent for the corresponding correlation function minus one, and to make $S(\xi)$ decay with increasing $\xi$ it requires $\eta>1$.

However, we find the above argument is oversimplified and thus does not generally hold. In TABLE~\ref{table:XYeta}, one could observe that generally $\eta'\neq\eta-1$ except for $Z$ and $XZX+YZY$. Actually, the value of $\eta'$ is extracted from calculation not in the continuum limit but within the discrete lattice, so 
\begin{equation}
S(\xi)= 2\sum_{r=1}^{\infty}C(r,\xi)+C(0,\xi).
\end{equation}
For $XX+YY$ at $p=0$, since $C(r,\xi=\infty)\sim (-1)^{r+1}r^{-2}$, for terms at large $r$ the sum reduces to
\begin{equation}
\sum_rC(r,\xi)\sim\sum_{\mathrm{odd}~r}\left[\frac{1}{r^2}-\frac{1}{(r+1)^2}\right]=\sum_{\mathrm{odd}~r}\frac{1}{r^2}\left[1-\left(1+\frac{1}{r}\right)^{-2}\right]\approx\sum_{\mathrm{odd}~r}\frac{2}{r^3}.
\end{equation}
Therefore effectively $\eta$ becomes 3 and thus the relation $\eta'=\eta-1$ still holds. For $XX-YY$ and $XY+YX$ at $p=\pi$, it could be argued similarly.

In fact, we found that $\eta'$ depends on the behavior of the correlation function calculated with MPS not only in the infrared limit but at all distances. For a MPS, $C(r,\xi)\sim r^{-\eta}e^{-r/\xi}$~\cite{PhysRevX.8.041033} is only true when $r$ is much larger than the lattice spacing, and thus there is a certain range $\Lambda<r<\xi'$ where $C(r,\xi)\sim r^{-\eta}$. At short distance $r<\Lambda$, the behavior of $C(r,\xi)$ relies on microscopic details. At sufficiently large distance $r\gg\xi'$, we have $C(r,\xi)\sim e^{-r/\xi}$. As a result, weird cancellation can happen to give the final value of $\eta'$ for $XZY-YZX$, $XZX-YZY$, and $XZY+YZX$, and it does not follow the previous simple argument.

\begin{table}
\centering
\begin{tabular}{ c|c|c|c }
\hline\hline
 $p$ & $G$ & $\eta'$ & $\eta$\\ 
\hline\hline
                    & $Z$      & ~~~~1.009~~~~ &~~~2~~~~\\\cline{2-4}
                    & $XX+YY$  & 1.985 & 2\\
~~~~0~~~~           & $XY-YX$  & 1.933 & 2\\\cline{2-4}
                    & $XZX+YZY$ & 1.008 & 2\\
                    & $XZY-YZX$ & 1.939 & 4\\
\hline
                    & $XX-YY$  & 2.005 & 2\\
~~~~$\pi$~~~~       & $XY+YX$  & 2.008 & 2\\\cline{2-4}
                    & $XZX-YZY$ & 2.046 & 4\\
                    & $XZY+YZX$ & 2.063 & 4\\                    
\hline\hline
\end{tabular}
\caption{The critical exponents $\eta$ in the correlation function $C(r,\xi=\infty)$ and the critical exponent $\eta'$ in $S(\xi)\sim\xi^{-\eta'}$ for the corresponding local conserved quantities in the spin-1/2 isotropic quantum $XY$ chain up to $N=3$. $\eta$ is first obtained from calculating correlation functions analytically by Jordan-Wigner transformation and Wick's theorem and then further checked numerically.}
\label{table:XYeta}
\end{table}

Notice that the correlation length manifested in the correlation function calculated with MPS can be different from the correlation length obtained by the formula $\xi=-\frac{1}{\mathrm{log}|\lambda_2/\lambda_1|}$, where $\lambda_i$ is the $i$th largest eigenvalue of the MPS transfer matrix $T=\sum_i\lambda_i|\lambda_i)(\lambda_i|$. Define
\begin{equation}
(G|=
\begin{diagram}
\draw (1,1) edge[out=180,in=180] (1,4);
\draw[rounded corners] (1,3.5) rectangle (2,4.5);
\draw[rounded corners] (1,0.5) rectangle (2,1.5);
\draw[rounded corners] (3,3.5) rectangle (4,4.5);
\draw[rounded corners] (3,0.5) rectangle (4,1.5);
\draw[rounded corners] (1,2) rectangle (4,3);
\draw (2.5,2.5) node (X) {$G$};
\draw (2.5,1.7) node {...};\draw (2.5,3.3) node {...};
\draw (1.5,1) node {$\bar{A}_L$};
\draw (1.5,4) node {$A_L$};
\draw (3.5,1) node {$\bar{A}_L$};
\draw (3.5,4) node {$A_L$};
\draw (1.5,1.5) -- (1.5,2);
\draw (1.5,3) -- (1.5,3.5);
\draw (3.5,1.5) -- (3.5,2);
\draw (3.5,3) -- (3.5,3.5);
\draw (2,4) -- (3,4); \draw (2,1) -- (3,1); 
\draw (4,4) -- (5,4); \draw (4,1) -- (5,1); 
\end{diagram}.
\end{equation}
The reason is that $(G|\lambda_i)$ can be zero for the first several $\lambda_i$'s, depending on the charge $G$ carries, and thus the correlation length will be instead given by $\xi=-\frac{1}{\mathrm{log}|\lambda_j/\lambda_1|}$, where $\lambda_j$ is the largest eigenvalue of $T$ that satisfies $(G|\lambda_j)\neq 0$. Nevertheless, we find that $\lambda_i=\lambda_2^{a_i}$ for $i>2$, which will only bring a constant prefactor $a_i$ to the correlation length and therefore will not affect the scaling.

\section{More results of the spin-1/2 $J$-$Q$ model}
At the transition point $Q/J\approx 0.84831$, we first use VUMPS with 1-site unit cell to obtain the ground state for various bond dimensions from $\chi=10$ to $\chi=400$ until the gradient converges to $10^{-12}$ and then apply our algorithm to these uniform MPS. When doing VUMPS, we only require the MPS to be real so as to impose time-reversal symmetry but not impose any other microscopic symmetries. When applying our algorithm to the MPS obtained by VUMPS from $N=2$, we impose the time-reversal, parity, and spin-flip symmetries as mentioned in the previous section. Since the one-site unit cell enforces translational invariance, we would get a (non-injective) equal weight superposition of all translational symmetry broken MPS ground state approximation, if it energetically favors a translational symmetry broken ground state within finite bond dimension, and thus it would limit the precision VUMPS can reach~\cite{PhysRevB.97.045145}. To avoid this, we perform a sublattice spin rotation around the $z$-axis by angle $\pi$ when doing VUMPS, which is important for it to converge. Due to this sublattice rotation, the $x$ and $y$ components of the generators move to $p=\pi$.

\begin{figure}[h!]
    \centering
    \includegraphics[scale=0.307]{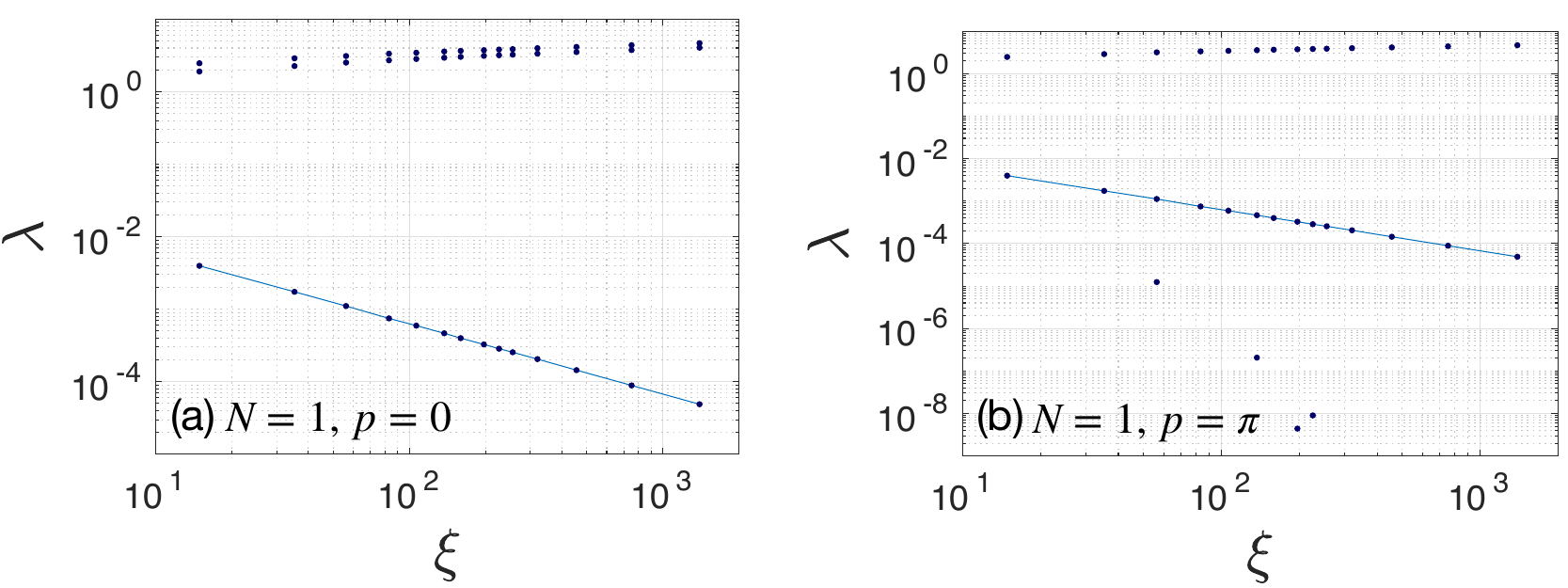}
    \caption{Log-log plots of the non-trivial eigenvalue spectrum (without imposing microscopic symmetries) of $\mathcal{F}$ versus the correlation length $\xi$ for the spin-1/2 $J$-$Q$ chain at $Q/J\approx 0.84831$ when $N=1$.}
    \label{fig:eigJQ}
\end{figure}

As a supplemental, here we also show the data for $N=1$, where we obtain the three generators for the microscopic $\mathrm{SU}(2)$ symmetry. For $p=0$, we get only one decaying solution, $G=Z$ (blue curve in Fig.~\ref{fig:eigJQ}(a)); for $p=\pi$, we get two decaying solutions corresponding to the other microscopic $\mathrm{SU}(2)$ generators $X$ and $Y$ (blue curve and scattered points below it in Fig.~\ref{fig:eigJQ}(b)). Notice that most of those scattered points are well below $10^{-8}$ and not shown.

\begin{figure}[h!]
    \centering
    \includegraphics[scale=0.305]{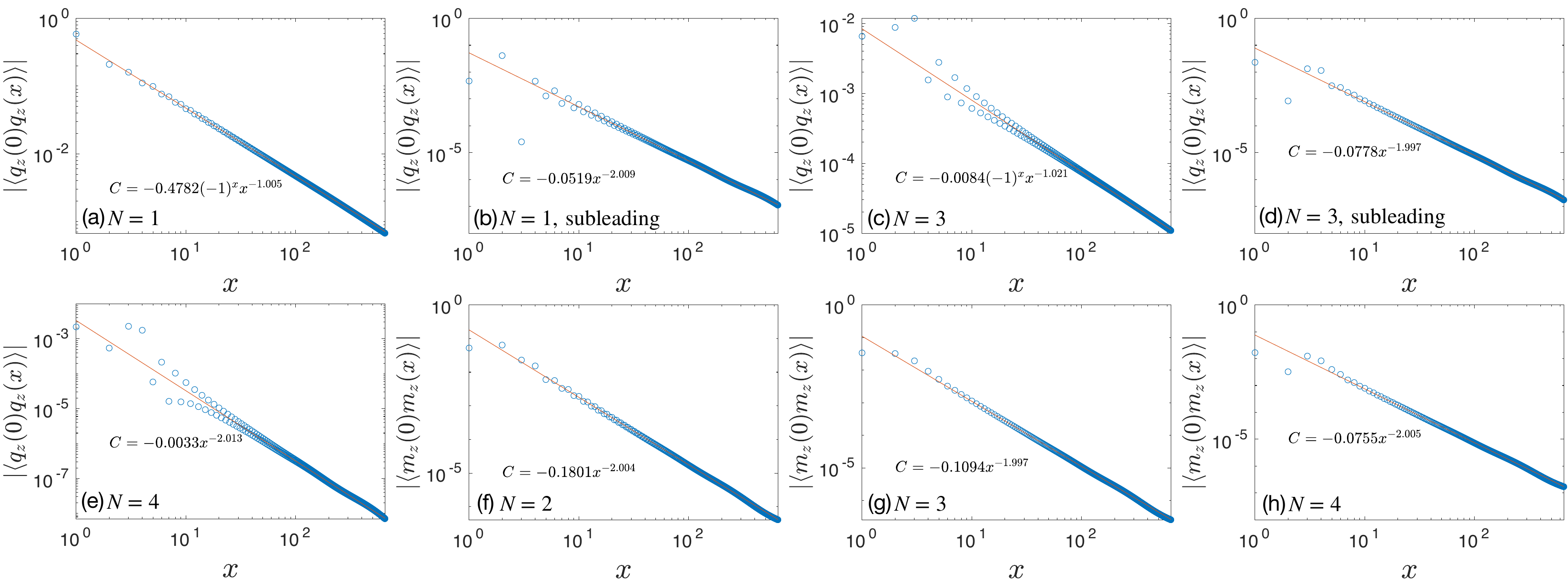}
    \caption{The absolute value of the correlation function $\langle q_z(0)q_z(x)\rangle$ and $\langle m_z(0)m_z(x)\rangle$ measured with MPS of bond dimension $\chi=240$, where $q_z$ and $m_z$ are the optimized solution for MPS of bond dimension $\chi=400$. The subleading part of the correlation function of the $N$-site operator $G$ is obtained by measuring the correlation function of $G\otimes I+I\otimes G$, and therefore the subleading part of the correlation function of the $1$-site $q_z$ is equivalent to the correlation function of the $2$-site $q_z$.}
    \label{fig:corrJQ}
\end{figure}

The conservation of currents in $1+1$d implies their scaling dimension must be 1~\cite{PhysRevLett.55.1355}, which means their correlation functions should decay as $r^{-2}$. This is indeed the case as shown in Fig.~\ref{fig:corrJQ}, which plots the correlation function of $q_z$ and $m_z$ for different $N$, where $Q_z=\sum_x q_z(x)$ and $M_z=\sum_x m_z(x)$ are the zero modes of the Kac-Moody generators. We can see that
\begin{equation}
    \langle q_z(0)q_z(r)\rangle\sim c_1(-1)^rr^{-1}+c_2r^{-2},~~\langle m_z(0)m_z(r)\rangle\sim r^{-2}.
\end{equation}
Notice that $\langle q_z(0)q_z(r)\rangle\sim r^{-2}$ for even $N$ because of the cancellation of the staggered part due to terms like $G\otimes I+ I\otimes G$, where $G$ is a $N-1$-site operator, as listed in Table~\ref{table:JQg}. This cancellation can be explained as follows. If $\langle G(0)G(r)\rangle\sim c_1(-1)^rr^{-1}+c_2r^{-2}$, then 
\begin{equation}
\begin{split}
&\langle (G\otimes I+ I\otimes G)(0)(G\otimes I+ I\otimes G)(r)\rangle\\
=&\langle G(0)G(r)\rangle+\langle G(0)G(r+1)\rangle+\langle G(1)G(r)\rangle+\langle G(1)G(r+1)\rangle\\
=&2[c_1(-1)^rr^{-1}+c_2r^{-2}]+c_1(-1)^{r+1}(r+1)^{-1}+c_2(r+1)^{-2}+c_1(-1)^{r-1}(r-1)^{-1}+c_2(r-1)^{-2}\\
\sim &4c_2r^{-2}.
\end{split}
\end{equation}

\begin{figure}[h!]
    \centering
    \includegraphics[scale=0.305]{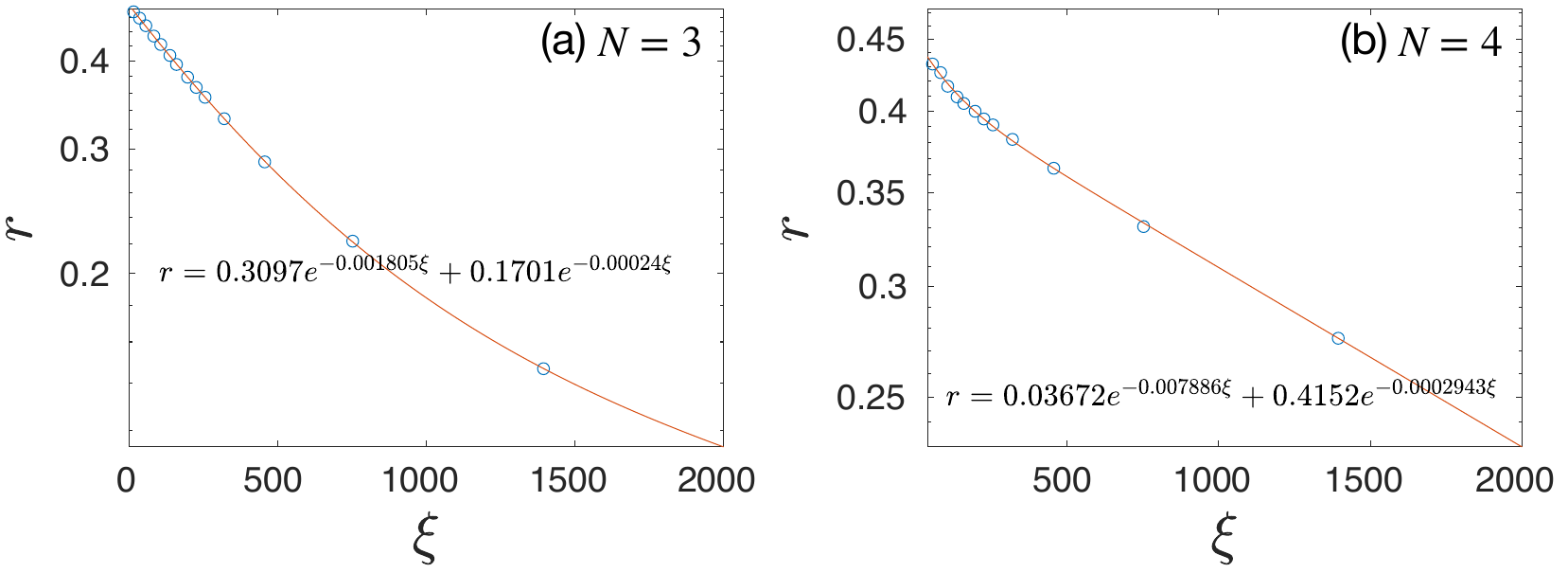}
    \caption{(a) The weight of the part excluding $S^z$ in $q_z$ for (a) $N=3$ and (b) $N=4$ versus the correlation length $\xi$.}
    \label{fig:resJQqz}
\end{figure}

Note that although for $q_{\alpha}$ at $N>2$ we got longer-range decoration to the 1-site spin operator $S_{\alpha}$ from terms in higher even-level Yangian generators. Those decorations will finally vanish as the correlation length goes to infinity, as shown in Fig.~\ref{fig:resJQqz}.

\begin{figure}[h!]
    \centering
    \includegraphics[scale=0.305]{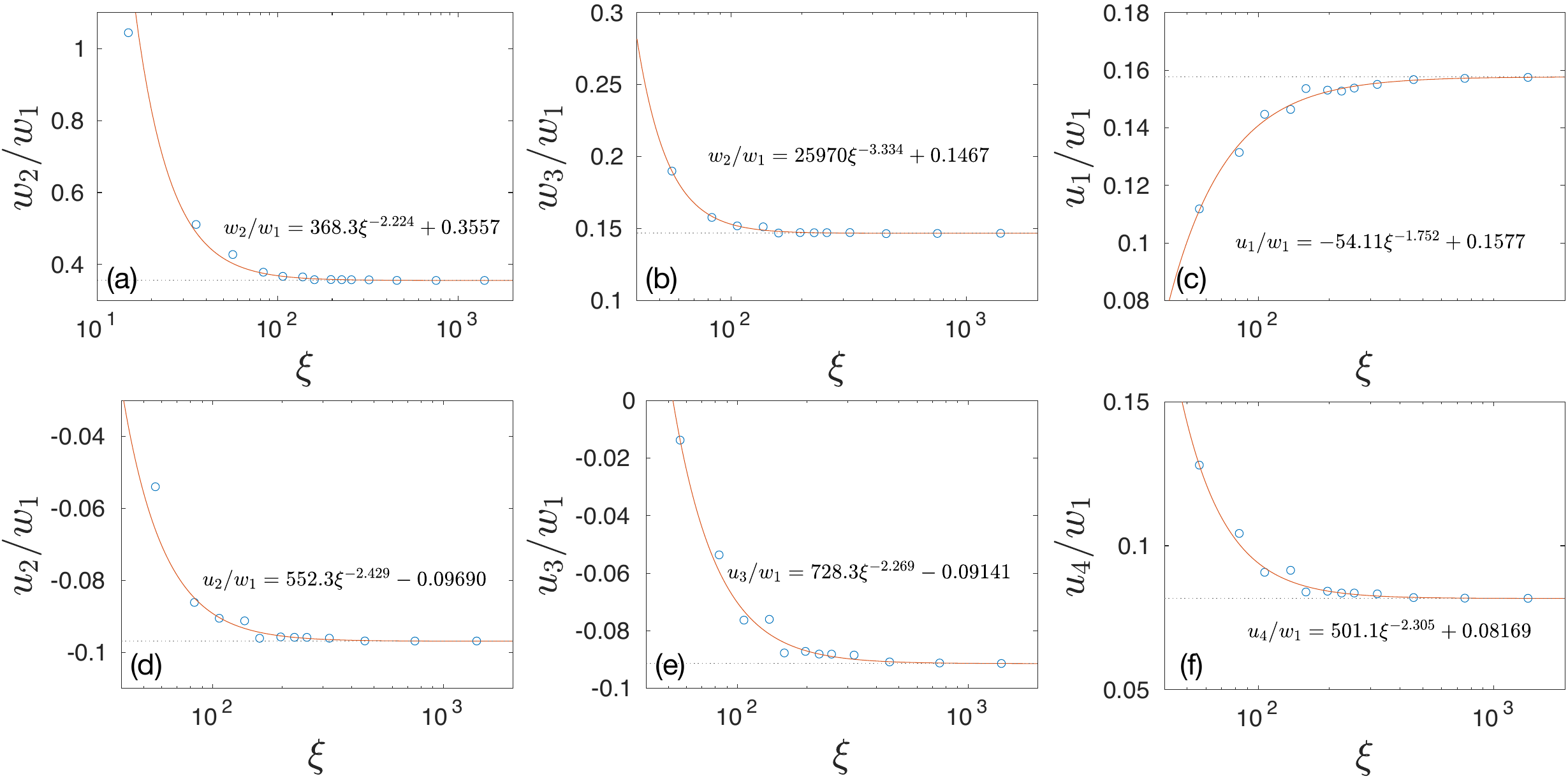}
    \caption{(a) The ratio between the coefficients of the terms in $m_z$ for $N=4$ versus $\xi$, similarly for $m_x$ and $m_y$.}
    \label{fig:ratioJQG2N4}
\end{figure}

In Fig.~\ref{fig:ratioJQG2N4} we also show the extrapolation of the coefficient ratio in $m_z$ for $N=4$.

\begin{table}[h!]
\centering
\scalebox{.82}
{
\begin{tabular}{ c|c|c|c|c }
\hline\hline
 $p$ & $N$ & $\mathcal{P}$ & $\mathcal{T}$ & $G$ \\ 
\hline\hline
                     & ~~~~1~~~~ &$+$&$-$           & $Z$        \\\cline{2-5}
                     & ~~~~2~~~~ &$+$&$-$           & $ZI+IZ$   \\\cline{3-5}
                     &           &$-$&$+$           & $XY-YX$        \\\cline{2-5}
                     &           &$+$&$-$            & $0.201918(ZII+IZI+IIZ)$  \\
                     & ~~~~3~~~~ & &            & $-0.016479(XXZ+ZXX+YYZ+ZYY)-0.028084(XZX+YZY)+0.004873ZZZ$\\\cline{3-5}
                     &           &$-$&$+$           & $0.168431[(XY-YX)I+I(XY-YX)]-0.075906(XIY-YIX)$ \\\cline{2-5}
                     &           &&           & $0.120163(IIIZ+IIZI+IZII+ZIII)$\\
                     &           &&           & $+0.020134(ZIZZ+ZZIZ)+0.019189(IZZZ+ZZZI)$\\
                     &           &$+$&$-$           & $-0.015721(IXXZ+IZXX+XXZI+ZXXI+IYYZ+IZYY+YYZI+ZYYI)$\\
 ~~~~0~~~~           &           &&           & $-0.012254(IXZX+XZXI+IYZY+YZYI)$\\
                     &           &&           & $+0.009580(XIXZ+ZXIX)+0.009581(YIYZ+ZYIY)-0.007155(XXIZ+ZIXX)-0.007156(YYIZ+ZIYY)$\\
                     &           &&           & $-0.003398(XIZX+XZIX)-0.003397(YIZY+YZIY)$\\\cline{3-5}
                     & ~~~~4~~~~ &&           & $0.090323(-IIXY+IIYX-IXYI+IYXI-XYII+YXII)$\\
                     &           &&           & $+0.048202(IXIY-IYIX+XIYI-YIXI)-0.039748(XIIY-YIIX)$\\
                     &           &$-$&$+$           & $-0.010666(XZZY-YZZX)+0.010015(XXXY-YXXX)+0.010014(XYYY-YYYX)$\\
                     &           &&           & $+0.006560(ZXYZ-ZYXZ)+0.005165(XXYX-YYXY-XYXX+YXYY)$\\
                     &           &&           & $+0.006189(XYZZ-ZZYX)-0.006190(YXZZ-ZZXY)$\\
                     &           &&           & $+0.005537(XZYZ-ZYZX)-0.005536(YZXZ-ZXZY)$\\\cline{3-5}
                     &           &$+$&$+$           & $0.0761[II(XX+YY-ZZ)+I(XX+YY-ZZ)I+(XX+YY-ZZ)II]$\\
                     &           &&           & $+0.034(XX+YY-ZZ)(XX+YY-ZZ)$\\
\hline
                     & ~~~~1~~~~ &$+$&$-$           & $(0.242768X+Y)$        \\\cline{5-5}
                     &           &&           & $(X-0.242768Y)$     \\\cline{2-5}
                     &           &$+$&$-$           & $0.242768(IX-XI)+(IY-YI)$    \\\cline{5-5}
                     & ~~~~2~~~~ &&           & $(IX-XI)-0.242768(IY-YI)$  \\\cline{3-5}
                     &           &$-$&$+$     & $0.242768(XZ+ZX)+(YZ+ZY)$\\\cline{5-5}
                     &           &&           & $(XZ+ZX)-0.242768(YZ+ZY)$\\\cline{2-5}
                     &           &&           & $0.242768(XII-IXI+IIX)+(YII-IYI+IIY)$ \\
                     &           &&           &$+0.139087(XYX+ZYZ)-0.081610(XXY+YXX-YZZ-ZZY)$ \\
                     &           &$+$&$-$           &$+0.033766(YXY+ZXZ)-0.019812(XYY+YYX-XZZ-ZZX)$\\
                     &           &&           &$-0.024133YYY-0.005859XXX$\\\cline{5-5}
                     & ~~~~3~~~~ &&           & $(XII-IXI+IIX)-0.242768(YII-IYI+IIY)$  \\\cline{3-5}
                     &           &&           & $0.242768[I(XZ+ZX)-(XZ+ZX)I]-0.109859(XIZ-ZIX)$ \\
                     &           &&           &$[I(YZ+ZY)-(YZ+ZY)I]-0.452528(YIZ-ZIY)$\\\cline{5-5}
                     &           &$-$&$+$           & $[I(XZ+ZX)-(XZ+ZX)I]-0.450663(XIZ-ZIX)$ \\
~~~~$\pi$~~~~        &           &&         &$-0.242768[I(YZ+ZY)-(YZ+ZY)I]+0.109407(YIZ-ZIY)$ \\
                     &           &&           &$-0.007000(XXZ-ZXX)-0.007346(YYZ-ZYY)$  \\\cline{2-5}
                     &           &&           &$0.242768(IIIX-IIXI+IXII-XIII)+(IIIY-IIYI+IYII-YIII)$\\
                     &           &&           &$+0.167554(YIYY-YYIY)+0.159692(YYYI-IYYY)$\\
                     &           &&           &$+0.130831(YXXI-IXXY+XXYI-IYXX)+0.130834(IYZZ-ZZYI+IZZY-YZZI)$\\
                     &           &&           &$+0.101976(IXYX-XYXI+IZYZ-ZYZI)+0.079728(XIXY-YXIX)-0.079730(YZIZ-ZIZY)$\\
                     &           &&           &$-0.059546(XXIY-YIXX)-0.059552(YIZZ-ZZIY)$\\
                     &           &$+$&$-$           &$+0.040675(XIXX-XXIX)-0.038766(IXXX-XXXI)$\\
                     &           &&           &$+0.031763(-IXYY+XYYI+YYXI-IYYX)+0.031762(IXZZ-XZZI-ZZXI+IZZX)$\\
                     &           &&           &$+0.028275(XIYX-XYIX)-0.028272(ZIYZ-ZYIZ)$\\
                     &           &&           &$+0.024755(IYXY-YXYI)+0.024756(IZXZ-ZXZI)$\\
                     &           &&           &$-0.019357(XYIY-YIYX)-0.019356(XZIZ-ZIZX)$\\
                     &           &&           & $+0.014457(XIYY-YYIX-XIZZ+ZZIX)-0.006866(YXIY-YIXY)-0.006863(ZIXZ-ZXIZ)$\\\cline{5-5}
                     &           &&           &$(IIIX-IIXI+IXII-XIII)-0.242768(IIIY-IIYI+IYII-YIII)$\\\cline{3-5}                  
                     &           &&           &$0.242768(IIXZ-IXZI+XZII+IIZX-IZXI+ZXII)$\\
                     &           &&           &$-0.251227(IXIZ-XIZI-IZIX+ZIXI)+0.271729(XIIZ+ZIIX)$\\
                     &           &&           &$+(IIYZ-IYZI+YZII+IIZY-IZYI+ZYII)$\\
                     &           &&           &$-1.034842(IYIZ-YIZI-IZIY+ZIYI)+1.119294(YIIZ+ZIIY)$\\
                     &           &&           &$-0.501097(YYYZ+ZYYY)+0.501096(YZZZ+ZZZY)$\\
                     & ~~~~4~~~~ &&           &$-0.256485(XYXZ+ZXYX-XZXY-YXZX)$\\
                     &           &&           &$-0.233853(XXYZ+ZYXX)-0.233854(XXZY+YZXX)$\\
                     &           &&           &$-0.121652(XXXZ+ZXXX)+0.121650(XZZZ+ZZZX)$\\
                     &           &&           &$+0.107080(YYZY+YZYY-ZYZZ-ZZYZ)$\\
                     &           &&           &$+0.084445(XYZX+XZYX))+0.062267(XYZY+YZYX)-0.062265(ZYXY+YXYZ)$\\
                     &           &&           &$-0.056771(XZYY+YYXZ+YYZX+ZXYY)$\\
                     &           &$-$&$+$           &$-0.025996(ZXZZ+ZZXZ)+0.025995(XXZX+XZXX)$\\
                     &           &&           &$+0.20501(YXZY+YZXY)-0.010763(YXXZ+ZXXY)$\\\cline{5-5}
                     &           &&           &$IIXZ-IXZI+XZII+IIZX-IZXI+ZXII$\\
                     &           &&           &$-0.533663(IXIZ+ZIXI-IZIX-XIZI)+0.440072(XIIZ+ZIIX)$\\
                     &           &&           &$-0.242768(IIYZ-IYZI+YZII+IIZY-IZYI+ZYII)$\\
                     &           &&           &$+0.129556(IYIZ+ZIYI-IZIY-YIZI)-0.106835(YIIZ+ZIIY)$\\
                     &           &&           &$-0.110877(XXXZ+ZXXX)-0.118095(XYYZ+ZYYX)+0.110874(XZZZ+ZZZX)$\\
                     &           &&           &$-0.072633(YXZY+YZXY)+0.057187(XXZX+XZXX)-0.057189(ZXZZ+ZZXZ)$\\
                     &           &&           &$+0.068534(YYXZ+ZXYY)+0.068521(XZYY+YYZX)+0.061302(XYZY+YZYX)-0.061288(YXYZ+ZYXY)$\\
                     &           &&           &$+0.028678(YXXZ+ZXXY)+0.026910(YYYZ+ZYYY)-0.026917(YZZZ+ZZZY)$\\
                     &           &&           &$+0.017634(XYZX+XZYX)-0.016634(XXYZ+ZYXX)-0.016630(XXYZ+ZYXX)$\\
                     &           &&           &$+0.014886(XYXZ+ZXYX)-0.014881(XZXY+YXZX)-0.013884(YYZY+YZYY-ZYZZ-ZZYZ)$\\
\hline\hline
\end{tabular}
}
\caption{Optimal lattice operator approximation of the conserved currents for the extended symmetry in the spin-1/2 $J$-$Q$ chain at $Q/J\approx 0.84831$ up to $N=4$ with $\chi=400$. There are 3 generators associated to the exact microscopic $\mathrm{SU}(2)$ symmetry and the corresponding eigenvalues of $\mathcal{F}$ decays in a power law with the the correlation length $\xi$: $Z$ has $\eta'\approx 1.027$, while $0.242768X+Y$ and $X-0.242768Y$ have $\eta'\approx 1.028$. Notice the effect on the signs from the sublattice rotation. And also notice that different (approximate) symmetry generators can be linearly combined with each other to form an eigenvector. + (-) means parity $\mathcal{P}$ or time reversal $\mathcal{T}$ even (odd).}
\label{table:JQg} 
\end{table}
\clearpage

\section{Yangians in the Haldane-Shastry model}
In the last section, we mentioned that the lattice form of the Kac-Moody generators in the $J$-$Q$ model looks like local truncation to the Yangian generators, which are exact symmetry of the Haldane-Shastry model. In this section, we give the explicit lattice forms of the Yangian generators up to level 3.

Yangian is a Hopf algebra and its level-$n$ generators can be defined recursively starting from the level-0 and level-1 generators~\cite{Drinfeld:466366}. For $\mathrm{SU}(2)$, the level-0 generators are the usual global $\mathrm{SU}(2)$ generators, i.e. $Q_0^{\alpha}=\sum_iS_i^{\alpha}$. The level-1 generators are
\begin{equation}
Q_1^{\alpha}=\frac{1}{2}\sum_{i\neq j}w_{ij}\mathrm{i}\epsilon^{\alpha\beta\gamma}S_i^{\beta}S_j^{\gamma}
\end{equation}
with $w_{ij}=(z_i+z_j)/(z_i-z_j)$ and $z_j=\mathrm{exp}(2\pi\mathrm{i}j/L)$, where $L$ is the size of the system with periodic boundary condition. By utilizing the recursive relation
\begin{equation}
\mathrm{i}\epsilon^{\alpha\beta\gamma}[Q_1^{\gamma},Q_{n-1}^{\beta}]=2Q_n^{\alpha},
\end{equation}
it can be derived that
\begin{equation}
Q_2^{\alpha}=\frac{1}{12}(N-1)(N-2)Q_0^{\alpha}-\sum_{i\neq j\neq k}w_{ij}w_{kj}S_i^{\alpha}\mathbf{S}_j\cdot\mathbf{S}_k
\end{equation}
and
\begin{equation}
Q_3^{\alpha}=\frac{1}{6}(N-1)(N-2)Q_1^{\alpha}+\frac{1}{2}\sum_{i\neq j\neq l\neq m}w_{li}(w_{ml}-w_{mi})(w_{jl}-w_{ji})\mathrm{i}\epsilon^{\alpha\beta\gamma}S_i^{\beta}S_j^{\gamma}\mathbf{S}_l\cdot\mathbf{S}_m.
\end{equation}
We could see that the odd (even) level Yangian generators are odd (even) under the parity transformation.

\section{More results of the Jiang-Motrunich model}
At $K_{2x}=K_{2z}=1/2,~J_x=1,~J_z\approx 1.4645$, we perform VUMPS~\cite{Note1} with 1-site unit cell from bond dimension $\chi=10$ to $\chi=400$ until the gradient converges to $10^{-12}$ and then apply our algorithm. We also tried two-site unit cell but did not see a notable difference in the results.

\begin{figure}[h!]
    \centering
    \includegraphics[scale=0.307]{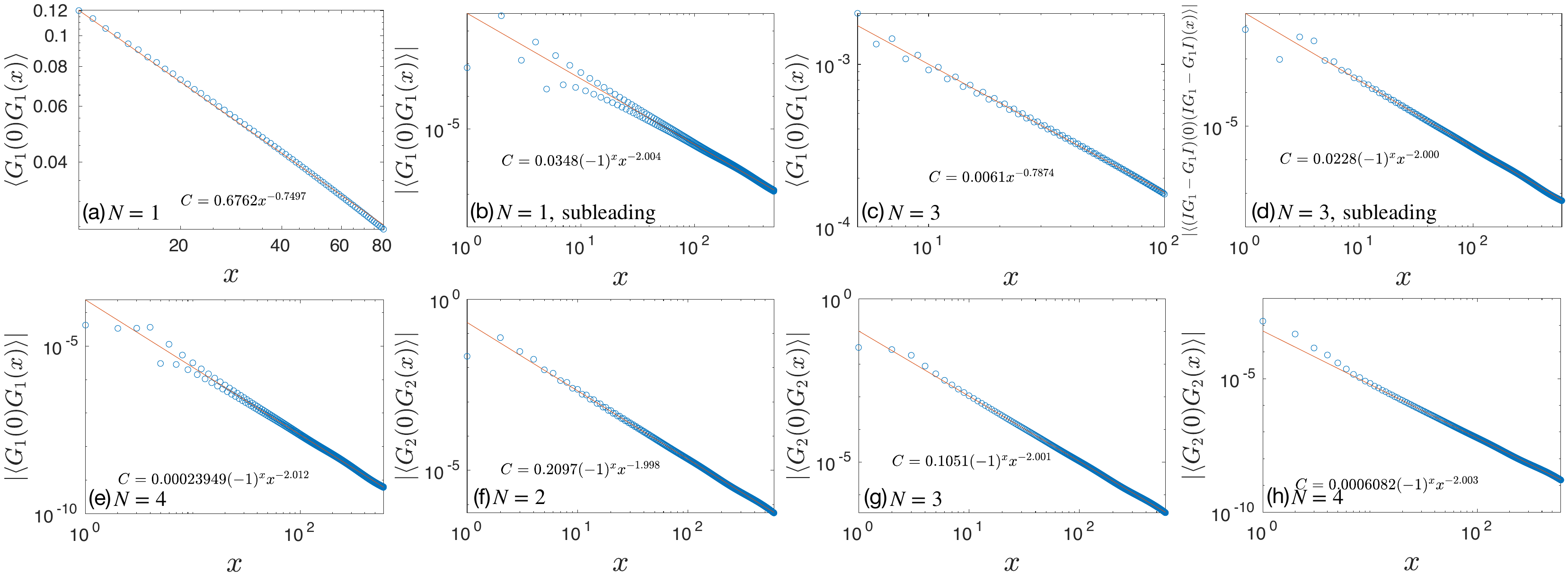}
    \caption{The absolute value of the correlation function $\langle G_1(0)G_1(x)\rangle$ and $\langle G_2(0)G_2(x)\rangle$ measured with MPS of bond dimension $\chi=240$, where $G_1$ and $G_2$ are the optimal solution for MPS of bond dimension $\chi=400$. The subleading part of the correlation function of the $N$-site operator $G$ is obtained by measuring the correlation function of $G\otimes I-I\otimes G$, and therefore the subleading part of the correlation function of the $1$-site $G_1$ is equivalent to the correlation function of the $2$-site $G_1$.}
    \label{fig:corrMotrunich}
\end{figure}

In Fig.~\ref{fig:corrMotrunich} we show the correlation function of $G_1$ and $G_2$ for different $N$ and confirm that their scaling dimension is pinned at one. We can see that
\begin{equation}
    \langle G_1(0)G_1(r)\rangle\sim c_1r^{-g/2}+c_2(-1)^rr^{-2},~~\langle G_2(0)G_2(r)\rangle\sim (-1)^rr^{-2},
\end{equation}
where $g\approx 1.5$ is the Luttinger parameter~\cite{PhysRevB.100.125137}. Notice that $\langle G_1(0)G_1(r)\rangle\sim (-1)^rr^{-2}$ for even $N$ because of the cancellation of the non-staggered part due to terms like $G\otimes I-I\otimes G$, where $G$ is a $N-1$-site operator, as listed in Table~\ref{table:Motrunichg}. This cancellation can be explained as follows. If $\langle G(0)G(r)\rangle\sim c_1r^{-1}+c_2(-1)^rr^{-2}$, then 
\begin{equation}
\begin{split}
&\langle (G\otimes I-I\otimes G)(0)(G\otimes I-I\otimes G)(r)\rangle\\
=&\langle G(0)G(r)\rangle-\langle G(0)G(r+1)\rangle-\langle G(1)G(r)\rangle+\langle G(1)G(r+1)\rangle\\
=&2[c_1r^{-1}+c_2(-1)^rr^{-2}]-c_1(r+1)^{-1}-c_2(-1)^{r+1}(r+1)^{-2}-c_1(r-1)^{-1}-c_2(-1)^{r-1}(r-1)^{-2}\\
\sim &4c_2(-1)^rr^{-2}.
\end{split}
\end{equation}

Here we also provide results of the eigenvalues of the optimization problem at $p=0$ from $N=1$ to $N=4$. None of them have scaling dimension one, and thus none of them are emergent conserved currents.
\begin{figure}[h]
    \centering
    \includegraphics[scale=0.307]{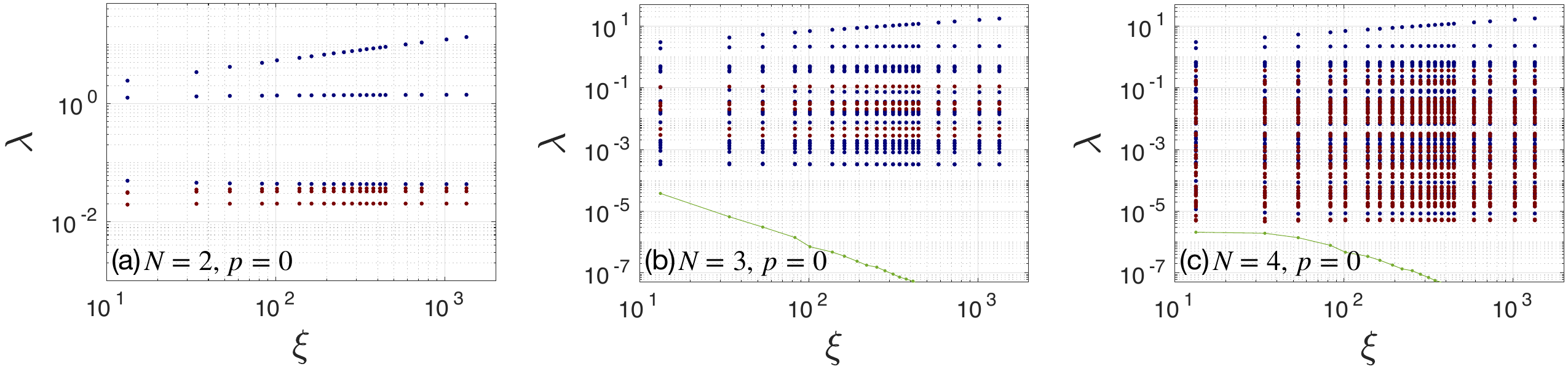}
    \caption{Log-log plot of the non-trivial eigenvalue spectrum of $\mathcal{F}$ (with time reversal and parity symmetry imposed) versus the correlation length $\xi$ at $p=0$ for the Jiang-Motrunich model at $J_z=1.4645$. The $G$'s associated with the blue (red) dots are parity even (odd) and time reversal odd (even). The green curve is the Hamiltonian.}
    \label{fig:eigMotrunich}
\end{figure}

In Table~\ref{table:Motrunichg} we list the optimal $G$'s we obtained with MPS of bond dimension $\chi=400$ from $N=1$ to $N=4$ for the Jiang-Motrunich model.
\begin{table}[h]
\centering
\scalebox{.85}
{
\begin{tabular}{ c|c|c|c|c }
\hline\hline
 $p$ & $N$ & $\mathcal{P}$ & $\mathcal{T}$ & $G$ \\ 
\hline\hline
                     & ~~~~1~~~~ & &          & ---   \\\cline{2-5}
~~~~0~~~~            & ~~~~2~~~~ & &          & ---   \\\cline{2-5}
                     & ~~~~3~~~~ &$+$&$+$         & $-(IXX+XXI)-1.4645(IZZ+ZZI)+(XIX+ZIZ)$   \\
\hline
                     & ~~~~1~~~~ &$+$&$-$         & $Z$        \\\cline{2-5}
                     & ~~~~2~~~~ &$-$&$+$         & $XY+YX$ \\\cline{3-5}
                     &           &$+$&$-$         & $ZI-IZ$    \\\cline{2-5}
                     &           &$-$&$+$         & $0.154265[(XY+YX)I-I(XY+YX)]+0.122085(XIY-YIX)$\\\cline{3-5}
                     & ~~~~3~~~~ &$+$&$-$         & $0.182771(-ZII+IZI-IIZ)$ \\
                     &           & &          & $+0.088579ZZZ+0.054162(YYZ+ZYY)+0.048393(XXZ+ZXX)+0.022477XZX-0.076692YZY$ \\\cline{2-5}
                     &           & &          & $II(XY+YX)-I(XY+YX)I+(XY+YX)II$\\
                     &           & &          & $+0.775400(-IXIY+IYIX+XIYI-YIXI)+0.178269(XIIY+YIIX)$\\
                     &           & &          & $+0.957965(XZZY+YZZX)-0.758349(XXXY+YXXX)+0.517924(XYYY+YYYX)$\\
 ~~~~$\pi$~~~~       &           &$-$&$+$          & $-0.397084(XZYX+ZYZX)$\\
                     &           & &          & $+0.334832(YZXZ+ZXZY)+0.333265(XYZZ+ZZYX)$\\
                     &           & &          & $+0.332877(XXYX+XYXX)+0.306209(YXZZ+ZZXY)$\\
                     & ~~~~4~~~~ & &          & $-0.222842(YXYY+YYXY)+0.188941(ZXYZ+ZYXZ)$\\\cline{3-5}
                     &           & &         & $0.015692(-IIIZ+IIZI-IZII+ZIII)$\\
                     &           & &          & $-0.094125(XIZX-XZIX)+0.072197(XXIZ-ZIXX)$\\
                     &           & &          & $-0.064930(XIXZ-ZXIX)+0.058821(ZIZZ-ZZIZ)$\\
                     &           &$+$&$-$          & $+0.048485(IXXZ+IZXX-XXZI-ZXXI)$\\
                     &           & &          & $+0.043120(IXZX-XZXI)+0.032214(YIZY-YZIY)-0.030060(IZZZ-ZZZI)$\\
                     &           & &          & $+0.012363(IYYZ-ZYYI+IZYY-YYZI)+0.011590(YYIZ-ZIYY)$\\
                     &           & &          & $-0.005822(YIYZ-ZYIY)+0.005798(IYZY-YZYI)$\\
\hline\hline
\end{tabular}
}
\caption{Optimal lattice operator approximation of the emergent conserved currents in the Jiang-Motrunich model at $J_z=1.4645$ up to $N=4$ with $\chi=400$. + (-) means parity $\mathcal{P}$ or time reversal $\mathcal{T}$ even (odd).}
\label{table:Motrunichg}
\end{table}

\end{document}